\documentstyle[12pt,fps]{article}
\makeatletter
\@addtoreset{equation}{section}
\makeatother

\newcommand{\bibi}{\bibitem}

\newcommand{\Lm}{\Lambda}
\newcommand{\un}{1\!\!1}

\newcommand{\al}{\alpha}

\newcommand{\bt}{\beta}
\newcommand{\lag}{\langle}
\newcommand{\rag}{\rangle}
\newcommand{\gm}{\gamma}
\newcommand{\pl}{\partial}
\newcommand{\Gm}{\Gamma}
\newcommand{\Sg}{\Sigma}
\newcommand{\dl}{\delta}
\newcommand{\ep}{\varepsilon}

\newcommand{\vep}{\varepsilon}

\newcommand{\zt}{\zeta}
\newcommand{\et}{\eta}

\newcommand{\kp}{\kappa}
\newcommand{\lm}{\lambda}

\newcommand{\yba}{\overline{y}}
\newcommand{\lmb}{\overline{\lambda}}
\newcommand{\vphibar}{\overline{\varphi}}

\newcommand{\rh}{\rho}
\newcommand{\sg}{\sigma}
\newcommand{\ta}{\tau}
\newcommand{\ph}{\phi}
\newcommand{\vr}{\varphi}

\newcommand{\ch}{\chi}
\newcommand{\ps}{\psi}

\newcommand{\khat}{\hat{k}}

\newcommand{\dlb}{\overline{\dl}}
\newcommand{\yt}{\tilde{y}}

\newcommand{\Sx}{\sum_x}

\newcommand{\rmv}{\protect{m_{\sg}/v_R}}
\newcommand{\av}[1]{\langle #1 \rangle}
\newcommand{\nnn}{\nonumber \\}
\newcommand{\htr}{\; {\textstyle \frac{1}{2} \mbox{Tr}\,}}

\newcommand{\Ps}{\Psi}

\newcommand{\Psb}{\overline{\Ps}}
\newcommand{\psb}{\overline{\ps}}

\newcommand{\dmu}{\partial_{\mu}}

\newcommand{\dsl}{\partial \!\!\!/}

\newcommand{\mh}{\hat{\mu}}

\newcommand{\chb}{\overline{\chi}}
\newcommand{\hmu}{\hat{\mu}}
\newcommand{\phat}{\hat{p}^2}
\newcommand{\rhh}{\hat{\rho}}
\newcommand{\aleq}{\mbox{}^{\textstyle <}_{\textstyle\sim}}
\newcommand{\ageq}{\mbox{}^{\textstyle >}_{\textstyle\sim}}
\newcommand{\half}{\mbox{{\small $\frac{1}{2}$}} }
\newcommand{\third}{\mbox{{\small $\frac{1}{3}$}} }
\newcommand{\quart}{\mbox{{\small $\frac{1}{4}$}} }
\newcommand{\eighth}{\mbox{{\small $\frac{1}{8}$}} }
\newcommand{\Sm}{\sum_{\mu}}
\newcommand{\Tr}{\mbox{Tr}}

\newcommand{\del}{\partial}
\newcommand{\dm}{\del_{\mu}}
\newcommand{\dg}{\dagger}
\newcommand{\ra}{\rightarrow}

\newcommand{\be}{\begin{equation}}
\newcommand{\ee}{\end{equation}}
\newcommand{\bea}{\begin{eqnarray}}
\newcommand{\eea}{\end{eqnarray}}
\newcommand{\eq}{\ref}
\newcommand{\beq}{\begin{equation}}
\newcommand{\eeq}{\end{equation}}
\newcommand{\cc}{\cite}
\newcommand{\lb}{\label}

\def \3{\ss}

\def\dateandnumber(#1)#2#3#4{
\vbox to 18mm{%
     \hbox to \textwidth{ \hspace*{14mm} \hsize=40mm%
            \vbox{%
                 \hbox to 40mm{\large #1 \hss}%
                 \hbox to 40mm{    \hss}%
                 \hbox to 40mm{    \hss}%
                 }%
                 \hss \hsize=80mm%
            \vbox{%
                 \hbox to 80mm{\hss \large #2}
                 \hbox to 80mm{\hss \large #3}
                 \hbox to 80mm{\hss \large #4}
                 }%
            \hspace*{14mm} }%
      \vss
    }
}
\def\titleofpreprint#1#2#3#4{{\LARGE \bf
\vbox to 43mm{%
     \vss
     \hbox to \textwidth{ \hspace*{14mm} \hsize=130mm%
            \hss \vbox{
                      \hbox to 130mm{\hss \LARGE \bf #1\hss}%
                      \hbox to 130mm{\hss \LARGE \bf #2\hss}%
                      \hbox to 130mm{\hss \LARGE \bf #3\hss}%
                      \hbox to 130mm{\hss \LARGE \bf #4\hss}%
                 }%
            \hss \hspace*{14mm} }%
      \vss
    }}
}
\def\listofauthors#1#2#3{{\large
\vbox to 22mm{%
     \vss
     \hbox to \textwidth{ \hspace*{14mm} \hsize=130mm%
            \hss \vbox{
                      \hbox to 130mm{\hss \large #1\hss}%
                      \hbox to 130mm{\hss \large #2\hss}%
                      \hbox to 130mm{\hss \large #3\hss}%
                 }%
            \hss \hspace*{14mm} }%
      \vss
    }}
}
\def\listofaddresses#1#2#3#4#5{{\small
\vbox to 18mm{%
     \vss
     \hbox to \textwidth{ \hspace*{14mm} \hsize=130mm%
            \hss \vbox{
                      \hbox to 130mm{\hss \small #1\hss}%
                      \hbox to 130mm{\hss \small #2\hss}%
                      \hbox to 130mm{\hss \small #3\hss}%
                      \hbox to 130mm{\hss \small #4\hss}%
                      \hbox to 130mm{\hss \small #5\hss}%
                 }%
            \hss \hspace*{14mm} }%
      \vss
    }}
}
\def\abstractofpreprint#1{{\normalsize
  \vbox to 110mm{%
     \vss
     \hbox to \textwidth{\hss \normalsize \bf Abstract \hss}%
     \vspace*{1cm} \normalsize
     #1
     \vss
    }}
}

\def\footnoteitem(#1)#2{
\begin{list}{#1}{\labelwidth4.0mm \leftmargin7.0mm
\labelsep2.5mm \rightmargin7.0mm \parsep0.5ex plus0.2ex minus0.1ex
\itemsep0ex plus0.2ex }
\item #2
\end{list}
}
\begin{document}
\dateandnumber(September 1992)%
{Amsterdam ITFA 92-23}%
{HLRZ J\"ulich 92-58}%
{UCSD/PTH 92-38}%
\titleofpreprint%
{      Can the Couplings in the                                }%
{      Fermion-Higgs Sector of the                             }%
{      Standard Model be Strong?                               }%
{                                                              }%
\listofauthors%
{Wolfgang Bock$^{1,\$}$, Christoph Frick $^{2,3,\#}$,          }%
{Jan Smit $^{1,\&}$ and Jeroen C. Vink $^{2,*,\times}$                }%
{                                                              }%
\listofaddresses%
{\em $^1$Institute of Theoretical Physics, University of Amsterdam,}%
{\em \\ Valckenierstraat 65, 1018 XE Amsterdam, The Netherlands }%
{\em $^2$HLRZ c/o KFA J\"ulich, P.O.Box 1913, 5170 J\"ulich, Germany}%
{\em $^3$Institute of Theoretical Physics E, RWTH Aachen,}%
{\em Sommerfeldstr., 5100 Aachen, Germany }%
\abstractofpreprint{
We present results for the renormalized quartic self-coupling $\lambda_R$
and the Yukawa coupling $y_R$ in a lattice fermion-Higgs model with
two SU(2)$_L$ doublets, mostly for large values of the bare couplings.
One-component (`reduced') staggered
fermions are used in a numerical simulation
with the Hybrid Monte Carlo algorithm.
The fermion and Higgs masses and the renormalized scalar field
expectation value are computed on $L^3 24$ lattices where $L$ ranges
from $6$ to $16$. In the scaling region these quantities are
found to have a $1/L^2$ dependence, which is used to determine
their values in the infinite volume limit. We then calculate the
$y_R$ and $\lambda_R$ from their tree level definitions in terms
of the masses and renormalized scalar field expectation value,
extrapolated to infinite volume. The
scalar field propagators can be described
for momenta up to the cut-off
by one fermion loop renormalized perturbation theory and
the results for $\lambda_R$ and $y_R$
come out to be close to the tree level unitarity bounds.
There are no signs that are in contradiction
with the triviality of the Yukawa and quartic self-coupling.
}
\noindent $\$$ e-mail: bock@phys.uva.nl   \\
\noindent $\#$ e-mail: hkf215@djukfa11.bitnet   \\
\noindent $\&$ e-mail: jsmit@phys.uva.nl \\
\noindent $*$  e-mail: vink@yukawa.ucsd.edu \\
\noindent $\times$ Present address: University of California, San Diego \\
\noindent \phantom{$\times$ Present address: }Department of
Physics, 9500 Gilman Drive 0319,\\
\noindent \phantom{$\times$ Present address: }La Jolla, CA 92093-0319, USA
\pagebreak
\section{Introduction}
The experiments in high energy physics are so far consistent
with the predictions of the perturbative Standard Model.
Two particles still await experimental
confirmation, the Higgs boson and the top quark.
The top quark seems to be on the verge of discovery.
Radiative corrections and high precision experiments
restrict the mass of the top quark to an interval of about
$120$-$180$ GeV \cite{TOP}, whereas the experiments of the CDF
collaboration give already
a lower bound of $91$ GeV with 95\% confidence level
\cc{CDF}. This indicates that the Yukawa
coupling of the top quark will be
relatively small and can still be treated
within perturbation theory.
Similar restrictions on the mass of the Higgs particle
are rather weak \cc{HIGGS}.
A lower bound on the Higgs boson mass was obtained in
ref.~\cc{Sh80},
where it was shown that it cannot be lighter than approximately
$50$-$100$ GeV, provided the mass of the top quark lies in the range
$100$-$200$ GeV. The determination of
upper bounds on the Higgs mass is a non-perturbative problem
and lies outside the scope of weak coupling expansions \cite{DaNe83}.
For recent analytical studies going beyond ordinary perturbation
theory using large $N$ techniques see ref.~\cite{LN}. \\

It is not excluded by the present status of the experiments
that there exists also a fourth fermion generation
with a neutrino heavier than half the $Z$ boson mass.
Since the systematics of the masses of the known fermions are unclear
until now there is no reliable extrapolation to the masses of
a fourth generation. Calculations of the radiative corrections
to the $\rho$ parameter indicate that
the mass split within the isospin doublets of a possible fourth generation
must be extremely small.
However, the constraints obtained by a
comparison of experimental
data with perturbative calculations may be doubtful if the
relevant couplings are large. \\

Experimentally it is important to know the largest possible
values of the Higgs boson and heavy fermion masses $m_H$ and
$m_F$ in the Standard Model. Theoretically this is an issue of
self-consistency of the model. Large ratios $m_H/v_R$ and
$m_F/v_R$ (where $v_R = 246$ GeV is the electroweak scale) imply
a large renormalized scalar self-coupling $\lm_R$ and Yukawa
coupling $y_R$. But it is well known that the Standard Model is
suspected to have the property called `triviality' and that
$\lm_R$ and $y_R$ can only increase if the regularization scale
$\Lambda$ decreases. For energies larger than $\Lambda$ the
theory is not consistent anymore and `new physics' has to take
over. If $m_H$ and/or $m_F$ get so large that $\Lambda$ gets
comparable in magnitude, the theory looses self-consistency and
new physics is noticeable at that mass scale. This scale is
somewhat fuzzy because the definition of $\Lambda$ depends on the
method of regularization and because it is not clear how large the
ratio $m_{H,F}/\Lambda$ can be before new physics takes over. \\

One can get an impression of the sensitivity to $\Lm$ by assuming
that the one-loop perturbative $\beta$-functions are valid for
momenta up to $\Lambda$ and identifying the bare couplings with
the running couplings evaluated at $\Lm$. Integration of the
renormalization group equations leads to a relation between the
bare couplings at the regulator scale
and renormalized couplings at the electroweak scale.
The renormalized couplings vanish logarithmically when
$\Lambda$ increases to infinity at
fixed bare couplings. Varying the bare couplings at fixed $\Lm$
the renormalized couplings go through an allowed region in the
coupling constant space. Examples for such allowed regions
in a fermion-Higgs model with two isospin doublets
and gauge couplings switched off are shown
in fig.~\ref{FLindn}, where we plotted the
ratio $m_{\sigma}/v_R=\sqrt{2 \lm_R}$ as a
function of $y_R=m_F/v_R$ (we shall use $m_H$ and $m_{\sg}$
interchangeably). The solid and dotted curves represent the
boundaries of such allowed regions for two different cut-off values.
The cut-off for the dotted curve is by a factor two larger
than for the solid curve. Such plots were first presented in
ref.~\cite{CaMa79} were $\Lm$ was chosen at the GUT scale, and
discussed further in ref.~\cite{Li86}. \\

The upper bound curve for $\lm_R$  does not depend very much on
$y_R$ and the lower bound curve for $\lm_R$
 increases rapidly when the value of $y_R$ is raised.
The upper (lower) bound curve corresponds to
infinite (zero) bare quartic coupling and a bare Yukawa coupling ranging
from zero to infinity.
Both curves join in a point where $y_R$  and $\lm_R$ have their largest
values.
When increasing $\Lm$ to infinity, the allowed region shrinks
to the origin
and its boundary depends only weakly on $\Lm$.
For smaller regularization scales the sensitivity to $\Lm$
increases. \\

The above exploration ignores of course the fact that infinite
bare couplings are outside the region of validity of the one-loop
$\beta$-functions. To overcome this we perform numerical simulations
using the lattice regularization with lattice distance $a$. The
regularization scale may be identified with the largest possible
momentum, which is $\pi/a$ for bosons and $\pi/2a$ for fermions,
or simply with the inverse lattice distance $1/a$.
The three values reflect the unavoidable arbitrariness in the
definition of $\Lm$. We shall use $\Lm = \pi/2a$
as the scale where the model breaks down and
`new physics' takes over.
In order to retain some scaling we shall limit the ratios
$m_{F,H}/\Lm \aleq 0.4$, which corresponds to $am_{F,H} \aleq 0.7$.
More precise definitions of upper bounds have
been studied in terms of limits to scaling violations
\cite{LuWe,HeNe92,GoKa92,Lang,Ku}. \\

It is reasonable first to investigate the question of upper bounds
within a simplified fermion-Higgs model and to ignore all couplings
in the electroweak theory which are small and can be treated well
within perturbation theory. For an overview about
the recent progress in this field using the lattice
regularization we refer to
ref.~\cc{REV}. \\

Using the lattice regularization for a
quantum field theory that involves fermionic field variables
one is confronted with the phenomenon of species doubling:
On a hypercubic lattice each fermion is accompanied by 15
doubler fermions with degenerate mass. This fermion doubling gives rise
to the following two problems:
a)
Eight of these species doublers couple with an opposite chiral charge
\cc{NiNi81} and spoil the chiral couplings
to the SU(2)$_L \otimes$U(1)$_Y$ gauge fields.
Several proposals have been developed to overcome this problem
\cc{REV}, but
so far none of them has been shown to succeed in formulating a
chiral gauge theory on the lattice.
Recently,
a thorough investigation of one of the proposals \cc{WY}
has raised strong doubts on its success
\cc{WYDEAD} although this is not generally accepted \cc{Ao}.
b) A second and more serious problem is caused by the large number
of mass degenerate fermion species.
A straightforward, non-chiral transcription
of a lattice fermion-Higgs model with one SU(2) doublet
on the lattice would lead
to a theory with 16 mass degenerate doublets, which is
not observed in nature.  It is therefore important
to develop methods which allow to reduce this
number of fermions. At the moment there exist two different
approaches, the mirror fermion method \cc{Mo87} and the staggered
fermion approach \cc{Sm88,Sm91,Sm92},
which enable us to reduce the number of isospin doublets to one.
In a numerical simulation with the Hybrid Monte Carlo algorithm (HMCA)
this number has to be doubled in both cases. \\

%
%
\begin{figure}
\centerline{
\fpsxsize=14.0cm
\fpsbox{FLindn.ps}
}
\vspace*{0.7cm}
\caption{ \noindent {\em The $m_{\sigma}$-$m_F$ plot for the fermion-Higgs
model with two mass degenerate doublets. The masses are given
in units of $v_R$.
The various points in this figure represent our non-perturbative
numerical results for the ratios $m_{\sg}/v_R$
and $m_F/v_R$ in the infinite volume limit.
The values for $av_R=\pi v_R/2\Lm $, given beside the frame
indicate the dependence of the cut-off $\Lm$.
The symbols correspond to those
in the phase diagram, fig.~2.
The vertical
and horizontal dashed lines give the tree level
unitarity bounds for the Yukawa and quartic self-coupling.
The solid
and dotted lines were obtained by integrating the one-loop
$\beta$-functions. The cut-off parameter for the dotted line is twice as
large as for the solid line.}}
\label{FLindn}
\end{figure}
In the mirror fermion model \cc{Mo87} each
fermion field is paired up with a mirror fermion field.
This `doubling' of the fermionic degrees of freedom allows for
chirally invariant Wilson mass terms to remove
species doublers of the original and the mirror
fermions from the spectrum \cc{FaKo91}.
The physical mirror fermion can also be
decoupled from the spectrum (in the fermion-Higgs model) by a proper
choice of the bare coupling parameters
and the model then describes one mass degenerate doublet in the scaling
region \cc{Mo92,FrLi92}. \\

In this paper we will follow a different proposal \cc{Sm91,Sm92} which is
based on the `reduced' or `real' staggered
fermion formalism \cc{ShTh81,DoSm83}.
The fermion-Higgs model we will investigate in this paper
has already been explored in ref.~\cc{BoSm92}, where we presented
preliminary results. The usual euclidean staggered fermions
on a four-dimensional hypercubic lattice describe four flavors
in the scaling region. By using the `reduced' staggered
formalism the number of staggered flavors can be reduced to two.
These two staggered flavors are coupled
to the Higgs field, leading to a model with only one doublet
in the scaling region. Since the staggered flavor symmetry group
is discrete, the full O(4) symmetry is broken by the Yukawa coupling
to a discrete subgroup and it is necessary
to add two scalar field counterterms
to recover the full symmetry in the scaling region \cc{BoSm92}.
We will not add these counterterms, but
present several analytical and numerical results
which show that the effects of the symmetry breaking are small in the
parameter region we are interested in. \\

The mirror fermion model and our model with
reduced staggered fermions are expected to reproduce in
the scaling region the target model, which
is described by the continuum action
\be
S_0=S_H+S_F  \lb{TARGET}
\ee
with
\bea
S_H &=& -\int d^4 x  \left[
\frac{1}{2}  \htr  \left\{
(\pl_{\mu} \phi )^{\dagger} (\pl^{\mu} \phi) \right \}
+ \frac{m_0^2}{2} \; \htr \left\{ \phi^{\dagger}
\phi \right \} + \frac{\lambda_0}{4}  \htr \left\{ \left(
\phi^{\dagger} \phi \right)^2 \right\}   \right] \lb{S0H}\\
S_F&=&  -\int d^4 x \left[
\sum_{j=1}^{N_D} \psb_j  \gm^{\mu} \partial_{\mu} \ps_j
+ y_0 \sum_{j=1}^{N_D}
            (\psb_{L,j} \ph \ps_{R,j} + \psb_{R,j} \ph^{\dg} \ps_{L,j})
\right]
           \;.\lb{S0F}
\eea
Here $\phi(x)$ is a 4-component scalar field in the $ 2 \times 2 $
matrix notation, $\ps_j(x)$ is an SU(2) doublet
and $N_D$ denotes the number of SU(2) doublets.
The bare coupling parameters $m_0^2$, $\lambda_0$ and $y_0$
are respectively the mass parameter, the quartic self-coupling
of the scalar field and the Yukawa coupling.  \\

We can regularize this model by introducing
a four-dimensional hypercubic lattice
with lattice spacing $a$ and replacing the derivatives $\partial_{\mu}
\phi(x)$ by $(\phi_{x+\hmu}-\phi_x)/a$.
It is convenient to  rescale the fields
\be
\phi (x) \ra \sqrt{2 \kp} \; \phi_x / a \;, \;\;\;\;\;\;\;
\ps (x) \ra \ps_x / a^{3/2}
                                  \lb{A}
\ee
and reparametrize the coupling parameters
\be
(a m_0)^2=\frac{1-2\lambda}{\kp} - 8\;, \;\;\;\;
\lambda_0   =\frac{\lambda}{\kp^2} \;, \;\;\;\;
y_0 = \frac{y}{ \sqrt{2 \kp} }\;.    \lb{BC}
\ee
As usual we shall mostly use lattice units, i.~e. $a=1$.
The actions (\eq{S0H}) and (\eq{S0F}) then get replaced by
\bea
S_H
&=&\sum_{x} \left\{ \kp \sum_{\mu}
\half \mbox{Tr} (\phi^{\dg}_x \phi_{x+ \mh}
                                 + \phi^{\dg}_{x+ \mh} \ph_x)
                 -\half \Tr \left[ \ph_x^{\dg} \ph_x
                 +\lm ( \ph_x^{\dg}\ph_x  - \un )^2  \right] \right\}
                  \lb{SH} \\
S_F&=&   -\sum_{x} \sum_{j=1}^{N_D} \left\{ \sum_{\mu} \half
         ( \psb_{x,j} \gm_{\mu} \psi_{x+\mh,j} -
           \psb_{x+\mh,j} \gm_{\mu}\psi_{x,j} )
   +    y (\psb_{x,L,j} \ph_x \ps_{x,R,j} + \psb_{x,R,j} \ph^{\dg}_x
             \ps_{x,L,j}) \right\}
               \;.  \nonumber \\
& & \lb{SF}
\eea
This model was extensively investigated in ref.~\cc{Search,BoDe91}.
Because of species doubling it
describes $16 \times
N_D$ doublets in the continuum limit (naive fermions). \\

In this paper we will use
the reduced staggered fermion method which
reduces this large number of SU(2) doublets to $N_D$. The reduced
staggered fermion formalism will be recalled in sect.~2.
For the study of the largest possible quartic coupling we consider
only $\lm=\infty$ which corresponds to
radially frozen Higgs fields, i.~e. $\ph^{\dg}_x\phi_x=\un$.\\

%
%
%
\begin{figure}
\centerline{
\fpsxsize=14.0cm
\fpsbox{Fphasd.ps}
}
\caption{ \noindent {\em Phase diagram of the
reduced staggered fermion model with $N_D=2$.
The solid lines indicate the position of the phase
transitions between the FM, PM, AM and FI phases.
At the points marked by the various symbols, we have performed numerical
simulations on a sequence of different lattices ranging in size from
$6^3 24$ to $16^3 24$. These symbols will be used throughout most of the
figures of this paper.
The dots represent some points at $\kp=0$ where we carried out
calculations only on a $12^3 24$ lattice.
}}
\label{Fphased}
\end{figure}
In the last part of this introduction we shall summarize the
main results.
The phase diagram for the model with $N_D=2$ obtained in
ref.~\cc{BoSm92} is shown in fig.~\ref{Fphased}.
There are four different phases in the $\kp$-$y$ plane:
A broken or ferromagnetic (FM), a symmetric
or paramagnetic (PM) phase, an antiferromagnetic
(AM) phase and a ferrimagnetic
(FI) phase. We will restrict ourselves in this paper to the
broken or ferromagnetic (FM) phase.
The AM and FI phases have to be regarded
most probably as lattice artefacts since their existence seems
to be related strongly to the hypercubic lattice
geometry. An interesting aspect of the phase diagram
is that the FM phase extends
into the negative $\kp$ region which is not accessible
in the continuum
parametrization (\eq{BC}).
We note that $y_0 \ra \infty$ for $\kp \searrow 0$
and according to eq.~(\eq{BC}) $y_0$ would become
imaginary for $\kp<0$. Our numerical results indicate that the whole
PM-FM phase transition line
falls into the same universality class and there is
no reason to ignore
the negative $\kp$ region. On the other hand
our quantitative results for the renormalized Yukawa coupling
suggest that the region with
$\kp<0$ may not be relevant since $y_R$ saturates when $\kp \searrow 0$
while keeping $v_R$ constant
and does not increase any more when $\kp$ is lowered beyond $\kp=0$.
The various symbols in fig.~\ref{Fphased} mark the points
in the phase diagram where we have carried out
the numerical simulation. The points are in three different
regions which are labeled by the Roman numerals (I), (II) and (III). \\

In fig.~\ref{FLindn} we have displayed
the Higgs mass $m_{\sg}$ as a function
of the fermion mass $m_F$.
The masses on the axis are given in units of the electroweak
scale $v_R\approx 246$GeV. The plot contains only our infinite volume
results for the ratios $m_{\sg}/v_R=\sqrt{2 \lm_R}$
and $m_F/v_R=y_R$.
The symbols in this figure label the position of the various points
in the phase diagram  and correspond to those in fig.~\ref{Fphased}.
The values of $a v_R$ listed beside the figure indicate
the cut-off values of the various points.
The arrows mark the tree level unitarity bounds for the
Yukawa and quartic self-coupling. The graph shows
that the numerical results are very close to these bounds
which implies that the renormalized couplings are not very strong.
Also our other results are in accordance with the triviality of the
Yukawa and quartic self-coupling.
It is remarkable that the solid curve, which was
obtained by integrating the one-loop $\bt$-function
is not very far  from the numerical results. \\

The outline of the paper is as follows: In sect.~2 we introduce the
model. Sect.~3 contains a perturbative
one-loop calculation for the Goldstone and Higgs particle propagators and
a discussion of the effects of the O(4) symmetry breaking. In sect.~4
we describe the numerical methods which we have
used for the determination of
the Higgs mass, the wave-function renormalization constant
of the Goldstone propagator and the fermion mass. Sect.~5 deals with
the extrapolation to infinite volume and discusses
the infinite volume results for the renormalized Yukawa and
quartic self-coupling. A summary of our results is given in sect.~6.
%
%
%
%
%
%
\section{The model}
We start this section by recalling the reduced staggered fermion method
which allows a reduction
of the number of 16 fermion species by a factor eight.
In the second part we will explain how the
two staggered flavors
can be coupled to the Higgs field. \\

Let us start from the naive euclidean
action
\be
S_K=  -\sum_{x \mu} \half
          ( \psb_{x} \gm_{\mu} \psi_{x+\mh} -
            \psb_{x+\mh} \gm_{\mu}\psi_{x} ) \;, \lb{NAIVE}
\ee
where the field $\psi_x$ is a usual four-component
Dirac spinor on the
lattice. Because of the fermion doubling phenomenon the action
(\eq{NAIVE}) leads to a model which can be
represented in the continuum by the following action
\be
S_K=  -\int d^4 x\sum_{j=1}^{N_F} \psb_j(x) \dsl \psi_j(x)
\;,\;\;\;\;\;N_F=16 \;,\lb{CON}
\ee
and it describes 16 massless fermions rather than one.
Using the usual staggered fermions this number
can be reduced by a factor four. This is achieved by performing a
spin-diagonalization transformation \cc{KaSm}
on the fields $\psi_x$ and $\psb_x$ in eq.~(\eq{NAIVE})
\be
\psi_x^{\al} = \sum_{\bt=1}^4 (\gm^x)^{\al \bt} \; \chi_x^{\bt} \;,\;\;\;
\psb_x^{\al} = \sum_{\bt=1}^4 \chb_x^{\bt} \; ((\gm^x)^{\dg})^{\bt \al} \;,
\lb{TT}
\ee
where $\gm^x \equiv  \gm_1^{x_1} \gm_2^{x_2} \gm_3^{x_3} \gm_4^{x_4}$.
After this transformation
the action (\eq{NAIVE}) is a sum of four identical terms
\be
S_{K} =   -\sum_{\al=1}^4 \sum_{x \mu} \half \et_{\mu x}
(\chb_x^{\al} \chi_{x+\mh}^{\al}
                 - \chb_{x+\mh}^{\al} \chi_x^{\al} ) \lb{LKS} \;.
\ee
The sign factor $\eta_{\mu x}$ in (\eq{LKS}) is given by
$\eta_{\mu x}=(-1)^{x_1+ \ldots + x_{\mu-1}}$ and the staggered fields
$\chi_x$ and $\chb_x$ are four-component
complex Grassmann variables. Using only one component and dropping the
superfix $\al$, it can be shown that the staggered fermion field
$\chi_x$ describes four Dirac fermions in the continuum limit.
The flavor and spin indices of the four staggered fermions
are spread out over the lattice and
do not appear in an explicit form.
To have control over the spin-flavor structure
it is convenient to introduce the $4 \times 4$ matrix fields \cc{Sm88}
\be
\Psi_x=\eighth \sum_{b}  \gm^{x+b}       \chi_{x+b} ,\;\;\;
\Psb_x=\eighth \sum_{b} (\gm^{x+b})^{\dg}\chb_{x+b} \lb{SDT} \;,
\ee
where in contrast to eq.~(\eq{TT})
we sum over the 16 corners of a unit lattice
hypercube, $b_{\mu}=0,1$ and the fields $\chi_x$ are one-component
complex Grassmann variables. The row (column)
matrix index of the $\Psi_x$ field
represents the spin (flavor) label and
vice versa for $\Psb_x$. Since the $\Psi$ field contains 16 times
as many degrees of freedom as the $\chi$ field,
not all components $\Psi^{\al\kp}$ are independent. Their
Fourier modes $\tilde{\Psi}(p)$ defined
in the restricted momentum interval
$-\pi/2<p_{\mu}\leq \pi/2$, however,
are independent \cc{Sm92}. The staggered fermion action
may now be written in the form
\be
  S_{K} =  -\sum_{x\mu} \half \Tr
               ( \Psb_x \gm_{\mu}\Psi_{x+\mh} -
                 \Psb_{x+\mh} \gm_{\mu}\Psi_{x} )\;. \lb{SFREE}
\ee
This form reduces in the classical continuum limit to
the action (\ref{CON}) with $N_F=4$. \\

This number of staggered flavors can be reduced once more by a factor two
by defining the $\chi_x$ fields on the odd sites
and the $\chb_x$ on the even sites of the hypercubic lattice, i.~e.
\be
\chi_x \ra \half (1-\ep_x) \chi_x \;,\;\;\;
\chb_x \ra \half (1+\ep_x) \chi_x
\ee
where $\ep_x=(-1)^{x_1+x_2+x_3+x_4}$. The insertion of these
relations into eq.~(\eq{SDT}) gives
\be
\Psi_x=\eighth \sum_{b} \gm^{x+b}
\half (1-\ep_{x+b}) \chi_{x+b} \;\;,\;\;\;
\Psb_x=\eighth \sum_{b} (\gm^{x+b})^{\dg}
\half (1+\ep_{x+b})  \chi_{x+b} \;.\lb{SDTR}
\ee
When inserting the relations (\eq{SDTR}) into
eq.~(\eq{SFREE}) one can reproduce the action for reduced (`real' or
`Majorana-like') staggered fermions \cc{DoSm83}
\be
S_{K} = -\half \sum_{x \mu} \et_{\mu x} \chi_x \chi_{x+\hmu}\;.
\lb{RST}
\ee
The restriction of the fields $\chi$ and $\chb$ to odd and even sites
corresponds to the projections
\be
\Psi \ra \half (\Psi - \gm_5 \Psi \gm_5) \;,\;\;\;
\Psb \ra \half (\Psb + \gm_5\Psb\gm_5) \;.
\ee
This implies the following structure for the matrix fields
\be
\Psb =    \left( \begin{array}{cc}
                     \psb_L &   0    \\
                        0   & \psb_R \end{array} \right),\;
      \Psi =    \left( \begin{array}{cc}
                        0    & \ps_R \\
                       \ps_L &    0    \end{array} \right) \;,
                       \lb{MATRIX}
\ee
where $\psb_L$, $\psb_R$, $\ps_L$ and $\ps_R$ are $2 \times 2$
matrices. The row and column
indices of the $\ps_L$ and $\ps_R$ fields
are respectively the Weyl-spinor and flavor labels, and
vice versa for $\psb_L$ and $\psb_R$. \\

The model in eq.~(\eq{RST}) is invariant under the
staggered fermion (SF) symmetry group which includes:\\
\noindent a) Shifts by one lattice distance
\be
\chi_x \ra \zt_{\rh x} \chi_{x+\rhh}\;,   \lb{SHIFT}
\ee
where $\zeta_{\mu x}=(-1)^{x_{\mu+1}+\cdots +x_4}$.\\
\noindent b) $90^{\circ}$ rotation
\be
\chi_x \ra S_R(R^{-1} x)  \chi_{R^{-1}x}\;,  \lb{RO}
\ee
where $R=R^{\sigma \rho}$ is the rotation
$x_{\rho} \ra x_{\sigma}$, $x_{\sigma} \ra -x_{\rho}$,
$x_{\tau} \ra x_{\tau}$ with $\tau \neq \rho, \sigma$ and
$S_R(x)=\half( 1+\eta_{\rho} \eta_{\sigma}
                -\zeta_{\rho}\zeta_{\sigma}
                +\eta_{\rho} \eta_{\sigma}
                \zeta_{\rho} \zeta_{\sigma})$. \\
\noindent c) Lattice parity
\be
\chi_x \ra (-1)^{ x_{1}+x_{2}+x_{3} }  \chi_{I x}\;,  \lb{AR}
\ee
where $I=I_s$ is the lattice parity transformation
$x_{4} \ra x_{4}$,
$x_{\tau} \ra -x_{\tau}$ for $\tau =1,2,3$.\\
\noindent d) Global U(1) symmetry
\be
\chi_x \ra e^{i \alpha \ep_x}  \chi_x\;, \lb{U1}
\ee
where $\alpha$ is a real phase. \\
The transformation (\eq{SHIFT}) can be interpreted as a
discrete flavor transformation \cc{DoSm83,GoSm84}.
The invariance of (\eq{RST}) under the symmetry
(\eq{U1}) implies fermion number conservation.
Next we shall couple the two reduced staggered
flavors to the Higgs field such that we recover in the scaling
region the target model of eq.~(\eq{TARGET}) with one SU(2) doublet.
We demand that the final form of the
action is invariant under the SF symmetry
group transformations since this ensures
the staggered flavor interpretation
in the scaling region. \\

In order to couple the reduced staggered fermion flavors to the
Higgs field we first introduce the $4\times 4$ matrix
\be
      \Phi =    \left( \begin{array}{cc}
                      0      & \ph      \\
                   \ph^{\dg} &  0      \end{array} \right) =
            - \Sm \vr_{\mu}\gm_{\mu}\;,
                 \lb{DEFPHI}
\ee
where $\vr_{\mu}$, $\mu=1,\ldots,4$ denote the
usual O(4) components of the Higgs fields. When using the relations
(\eq{MATRIX}) and (\eq{DEFPHI}) one can show that the action
\be
   S_F     =   -\sum_x \left[        \Sm \half \Tr
          ( \Psb_x \gm_{\mu}\Psi_{x+\mh} -
            \Psb_{x+\mh} \gm_{\mu}\Psi_{x} ) +
   y \Tr (  \Psb_x \Psi_x \Phi_x^T)          \right] \lb{SFM}
\ee
reduces in the classical continuum limit to the action of the
target model
in eq.~(\eq{S0F}) with $N_D=1$.
After inserting the transformations (\eq{SDTR}) into
eq.~(\eq{SFM}) the final form of the
fermionic action in terms of the $\chi$ fields reads
\be
 S_F = -\half  \sum_{x \mu} \chi_x \chi_{x+\hmu}
       ( \eta_{\mu x} + y \ep_x \zeta_{\mu x} \vphibar_{\mu x} )
       = -\half \sum_{x,y} \chi_x M_{xy} \chi_y \;, \lb{SCHI}
\ee
where
\be
\vphibar_{\mu  x} = \frac{1}{16} \sum_b \varphi_{\mu, x-b}
\lb{PHIHC}
\ee
is the average of the scalar field over a lattice hypercube.
The hypercubic Yukawa coupling arises naturally from the summation
over the corners of a lattice hypercube
which we introduced in the definition of the
$\Psi$ matrices in eq.~(\eq{SDTR}).
The fermion matrix $M$ in eq.~(\eq{SCHI}) is antisymmetric and real.
The final form of the lattice action at $\lm=\infty$ is then given by
\be
S=2 \kp \sum_{x \mu} \varphi_{\mu x} \varphi_{\mu,x+\hmu}
     -\half  \sum_{x \mu} \chi_x \chi_{x+\hmu}
     (\eta_{\mu x} + y \ep_x \zeta_{\mu x} \vphibar_{\mu x} ) \;,\lb{SS}
\ee
with $\sum_{\mu=1}^4 \varphi_{\mu x}^2 = 1$.
The action (\eq{SS}) is invariant under the SF symmetry group if the
Higgs field transforms in the following way under:\\
\noindent a) Shifts by one lattice distance:
\be
\varphi_{\mu  x} \ra  (1-2\delta_{\mu\rho}) \varphi_{\mu , x+\rhh}\;,
\lb{PS}
\ee
\noindent b) $90^{\circ}$ rotations:
\be
\varphi_{\mu   x} \ra R_{\mu \nu } \varphi_{\nu , R^{-1} (x+n)-n}  \;,
\lb{PR}
\ee
where $n=(\half,\half,\half,\half)$, \\
\noindent c) and lattice parity:
\be
\varphi_{\mu  x} \ra (2\delta_{\mu 4}-1)
\varphi_{\mu , I x} \;.
\lb{PA}
\ee
The  additional shifts by the vector
$n$ in eq.~(\eq{PR}) are needed because
the relation between the $\Psi$ and $\chi$ fields in eq.~(\eq{SDTR})
is not manifestly rotationally covariant.
The invariance under rotation would have been
more transparent if we would have associated
the $\Psi$ and $\varphi$ fields from the beginning
with the dual lattice which is shifted by the vector $n$ with respect to
the lattice for the $\chi$ fields.
The scalar field does not transform under the U(1)
symmetry (\ref{U1}).\\

The action is not invariant, however, under the full O(4)
flavor group,
but one expects to be able to recover this invariance
in the scaling region. In the scaling region operators
with dimension larger than four become irrelevant. There are, however,
two operators with dimension four which respect the discrete
symmetries (\eq{PS})-(\eq{PA}), but break O(4):
\be
O^{(1)}=  \sum_{x \mu} \varphi_{\mu  x}^4 , \;\;\;\;
O^{(2)}=  \frac{1}{2} \sum_{x \mu} (\varphi_{\mu , x+\hmu} -\varphi_{\mu  x})^2
. \lb{C2}
\ee
We will show in the next section that these terms are
indeed generated by the quantum fluctuations.
In order to recover the full O(4) symmetry one has to add
these operators as counterterms to the
action (\eq{SS})
\be
S \ra S + \vep_0 O^{(1)} + \dl_0 O^{(2)}\;,
\ee
and tune the coefficients
$\vep_0$ and $\dl_0$ as a function of the bare parameters $\kp$ and $y$
such that the O(4) invariance gets restored in the scaling region.
In this paper we will not add these counterterms, but
present several analytic and numerical results which show that
the effect of the symmetry breaking is small in the
parameter region of interest.
\section{One-loop fermion effects on the scalar propagator}
In the pure O(4) model the numerically measured scalar field
propagator has a momentum dependence that is nearly of the free
field form because the renormalized self-couplings are small.
The Yukawa interaction with the fermions affects the scalar
propagator in three ways: the masses and wave-functions are
renormalized, the additional self-energy gives rise to a
more complicated momentum dependence than the almost pure pole
found in the O(4) model, and with our staggered fermion method
the fermions induce O(4) symmetry breaking effects.
The last two effects can be studied in one fermion loop renormalized
perturbation theory. This is useful even at large bare couplings because
the renormalized couplings turn out to be relatively small.
By taking into account the one-loop
effects we can then more reliably extract the scalar masses and
wave-function renormalization constants from the numerical data
and estimate the effect of O(4) symmetry breaking on the
renormalized couplings. \\

If the fermion effects are neglected the
scalar sector of the model has the following approximate effective action,
\be
 S_{eff} \approx -\sum_x \left[  \frac{1}{2} \dm \vr_{R\nu}\dm \vr_{R\nu}
        + \frac{m_R^2}{2} \vr_{R\mu}\vr_{R\mu} + \frac{\lm_R}{4}
          (\vr_{R\mu}\vr_{R\mu})^2 \right] \;,\lb{SR}
\ee
where $\dmu \vr_x = \vr_{x+\hmu} - \vr_x$ and
the subscript $R$ denotes renormalized quantities.
In the broken phase we can decompose the scalar field in a
Higgs mode, $\sg$, and three Goldstone modes, $\pi^a$, $a=1,2,3$,
according to
\be
 \vr_{R\mu}= v_R e^4_{\mu} + \vr_{\mu}^{\prime}
=(v_R+\sigma_R) e^4_{\mu}+ \pi^a_R e^a_{\mu} \;.
\lb{DECOMP}
\ee
The $\{e^{\al}_{\mu}\}$ form an orthogonal set of O(4) unit
vectors and $v_R$ is the scalar field expectation value. Before
taking fermion effects into account, the choice of $e_{\mu}$'s
is arbitrary and after substituting the decomposition (\ref{DECOMP})
into the action (\ref{SR}) one finds the usual tree
level relations for the masses of the Higgs and Goldstone modes,
\be
m_{\sg}^2 = 2\lm_R v_R^2, \;\;\;
m_{\pi}^2 = 0,\;\;\;
v_R^2 = \frac{-m_R^2}{\lm_R},
\lb{TRHIGGS}
\ee
as well as the three point interactions
$  \lm_R v_R [\sg_R^3 + \sg_R \pi_R^a \pi_R^a ]$.
The scalar propagator
\be
G_{\mu\nu}(k) = \left\langle \frac{1}{V} \sum_{x,y}
\vr_{\mu}^{\prime} \vr_{\nu}^{\prime} \exp (ik(x-y)) \right\rangle
\lb{GGSS}
\ee
is
given by
\bea
G_{\mu\nu}^{-1}(k)&=& l_{\mu\nu}(\khat^2 + m_{\sg}^2) +
t_{\mu\nu} (\khat^2 + m_{\pi}^2),\;\;\;
\khat^2 = 2 \Sm (1-\cos k_{\mu}), \label{SCALPROP}\\
l_{\mu\nu}&=&e_{\mu}^4 e_{\nu}^4,\;\;\;
t_{\mu\nu}= e_{\mu}^a e_{\nu}^a,\;\;\;
l_{\mu\nu}+ t_{\mu\nu} = \dl_{\mu\nu},
\eea
where $l_{\mu\nu}$ and $t_{\mu\nu}$ are the longitudinal and
transverse projectors onto the $\sg$ and $\pi$ subspaces.\\

%
\begin{figure}
\centerline{
\fpsxsize=9.0cm
\fpsbox{Fdiag.ps}
}
\vspace{-5.0cm}
\caption{ \noindent {\em Feynman diagrams for the
one fermion loop contributions to the vacuum expectation value (a) and
to the scalar field self-energy (b and c) in the FM
phase.
}}
\label{Fdiag}
\end{figure}

To find the effect of the staggered fermions on the scalar propagator
we compute the one-loop diagrams shown in fig.~\ref{Fdiag}.
The Feynman rules for the fermions can be derived from the action
\be
    S_F  =  -\half  \sum_{x \mu} \chi_{x} \chi_{x+\hmu}
  ( \et_{\mu x} + m_F e^4_{\mu} \ep_x \zt_{\mu x})
          -\half y_R \sum_{x \mu} \chi_{x} \chi_{x+\hmu}
\vphibar_{\mu  x}^{\prime} \ep_x\zt_{\mu x}\;,
    \lb{SFR}
\ee
which results after inserting (\eq{DECOMP}) into eq.~(\eq{SCHI}).
The fermion mass in the first term is given by the usual tree
level relation
\be
       m_F = y_R v_R \;.    \lb{YR}
\ee
The bar on $\vr^{\prime}$ indicates hypercubic averaging, as in
(\ref{PHIHC}). \\

We shall use in the following
the staggered fermion formalism developed in
refs.~\cite{DoSm83,GoSm84}. The fermion propagator follows from the
first term in eq.~(\ref{SFR})
\bea
S_{AB}(p)\dlb(p-q) &=&
   \sum_{x,y} e^{-i(p+\pi_A)x}\av{\ch_{x}\ch_{y}} e^{i(q+\pi_B)y}
                    \nnn
  S_{AB}(p)& = & \frac{ \Sm[ -i\Gm_{\mu AB} \sin p_{\mu}
                   + m_Fe^4_{\mu}(\Xi_{\mu}\Xi_5\Gm_5)_{AB}
                      \cos p_{\mu}]          }
      {  \Sm [\sin^2 p_{\mu} + (e^4_{\mu})^2 m_F^2
                      \cos^2 p_{\mu}]      }\;,  \lb{SABREN}  \\
\dlb(p-q) &=& V\dl_{p,q}\;\; \mbox{mod}(2\pi)\;. \lb{DELBAR}
\eea
The lattice momentum of the fermion is in the restricted interval
$p_{\mu} \in (-\pi/2,\pi/2]$, $\pi_A$, $A=1,\ldots,16$, are the
momentum four-vectors with components equal
to $0$ or $\pi$ and $V$ is the lattice volume. The 16-dimensional
gamma and flavor matrices $\Gm_{\mu}$ and $\Xi_{\mu}$
form a Clifford algebra with $\{\Gm_{\mu},\Gm_{\nu}\}=2\dl_{\mu\nu}$,
$\{\Xi_{\mu},\Xi_{\nu}\}=2\dl_{\mu\nu}$,
$[\Gm_{\mu},\Xi_{\nu}]=0$.\\

The second term in (\ref{SFR}) contains the interaction with the
scalar fields averaged over a hypercube. The corresponding
vertex function is given by
\bea
 \Gm_{\mu AB}(p,-q,k) & = & -y_R h(k) e^{ik_{\mu}/2}
\cos(q- \half k_{\mu}) e^{i\pi_{B\mu}}
\dlb(p-q+k+\pi_A+\pi_B+\pi_{\ep}+\pi_{\zt_{\mu}}) \;,\nnn
h(k) &=& \frac{1}{16} \sum_b e^{-ikb}=
e^{-ikn} \prod_{\rh} \cos \frac{k_{\rh}}{2} \; ,
\eea
where $n=(\half,\half,\half,\half)$ is the vector introduced earlier in
(\ref{PR}). The quantities
$p+\pi_A$ and $q+\pi_B$ are the outgoing and incoming wave
vectors of the fermion field, with $p$ and $q$
in the restricted interval mentioned above,
and $k$ is the outgoing momentum of the scalar field
$\vr_{\mu}^{\prime}$. The $\pi_{\ep}$ and $\pi_{\zt_{\mu}}$ are defined such
that $\ep_x = \exp(i\pi_{\ep}x)$ and
$\zt_{\mu x}=\exp(i\pi_{\zt_{\mu}}x)$.
The factor $h(k)$ is due to the hypercubic fermion-scalar
coupling. In the classical continuum limit the momenta in lattice
units approach zero and
\be
 \Gm_{\mu}(p,-q,k)\ra -y_R \Xi_5\Gm_5\Xi_{\mu} \dlb(p-q+k).
\ee
Note that $y_R$ is a bare Yukawa coupling as
it does not contain at this stage the effects of the fermion
interactions.\\

Let us first consider the one loop fermion effect
on the vacuum expectation value of the scalar field (cf.
fig.~\ref{Fdiag}a):
\bea
\sum_x e^{-ikx} \langle \vr_{\mu x}\rangle &=& v_R e_{\mu}^4
\dlb(k) - \half N_D \sum_{\nu}
\left[ G_{\mu\nu}(k) \sum_p \frac{1}{16} \Tr
\{ \Gm_{\nu}(p,-p,k) S(p) \} \label{VEV} \right]\\
&=&\left[v_R e_{\mu}^4 + \sum_{\nu} \left( \frac{l_{\mu\nu}}{m_{\sg}^2}
+ \frac{t_{\mu\nu}}{m_{\pi}^2} \right) I_{\nu} \right]\dlb(k)\; ,\nnn
I_{\nu} &=& \half y_R m_F N_D e_{\nu}^4 \sum_p \frac{\cos^2 p_{\nu}}{
\sum_{\rh} [ \sin^2 p_{\rh} + m_F^2 (e_{\rh}^4)^2 \cos^2 p_{\rh} ]}\; ,
\eea
where $N_D$ is the number of fermion doublets (which is two in
our numerical work) and $\sum_p$ is a normalized sum over the
lattice momenta.
The direction of spontaneous symmetry breaking
$e_{\mu}^4$ is compatible with the fermion loop correction if the
latter has no transverse component, i.e. $\sum_{\nu} t_{\mu\nu} I_{\nu} =
0$.
This is the case for $e_{\mu}^4$ = $(\pm 1,0,0,0)$, \ldots,
$(0,0,0,\pm 1)$, $(\pm 1,\pm 1,0,0)/\sqrt{2}$, \ldots, $(\pm 1,\pm 1,\pm 1,
0)/\sqrt{3}$, \ldots, $(\pm 1,\pm 1,\pm 1,\pm 1)/2$.
By studying the one fermion loop effective potential
\be
V_{eff}(\vr) = \frac{N_D}{2} \sum_p \frac{1}{16} \Tr \ln
\frac{\Sm[
-i\Gm_{\mu} \sin p_{\mu} + y_R \vr_{R\mu}\Xi_{\mu}\Xi_5\Gm_5 \cos p_{\mu}]
}{\Sm[
\sin^2 p_{\mu} + y_R^2 \vr_{R\mu}^2 \cos^2
p_{\mu}]}\label{EFFERPOT}
\ee
for small $y_R$ we found in ref.~\cite{BoSm92} that only
$e_{\mu}^4 = (\pm 1,\pm 1,\pm 1,\pm 1)/2$ are local minima which
correspond to a ground state; the others are saddle points or
local maxima. We checked this conclusion by a numerical study of
the $\langle \vr_{\mu }\rangle$ probability distribution. In the
following we shall understand $e_{\mu}^4$ to be one of the 16
unit vectors
\be
e_{\mu}^4 = \frac{1}{2} (\pm 1,\pm 1,\pm 1,\pm 1) \label{E4MU}.
\ee

Note that the hypercubic coupling prohibits a fermion induced
staggered magnetization $\langle \vr_{\mu x} \rangle \propto
\ep_x$. One way to see this is that the factor $h(k)$ in
$\Gm_{\mu}(p,-p,k)$ in (\ref{VEV}) vanishes when $k_{\mu} =
\pi$.\\

We now turn to the fermion contribution to the scalar
self-energy corresponding to the diagrams b and c shown in fig.~\ref{Fdiag}.
The tadpole contribution of fig.~\ref{Fdiag}b is given by the
following expression:
\bea
\Sg_{\mu\nu}^{(b)}(k) &=& \half y_R^2 N_D (3 l_{\mu\nu} + t_{\mu\nu})
\sum_p \frac{\cos^2 p_1}{D(p)}\; , \label{TADPO}\\
D(p)&=&\sum_{\rh} \left[ \sin^2 p_{\rh} + \frac{1}{4} m_F^2 \cos^2 p_{\rh}
\right]  \; , \label{DDEF}
\eea
which is independent of $k$.
Diagram c in fig.~\ref{Fdiag} leads to the contribution
\bea
\Sg_{\mu\nu}^{(c)}(k) &=& \half y_R^2 N_D |h(k)|^2
e^{i\half (k_{\mu} - k_{\nu})} \sum_p
\cos (p+\half k)_{\mu}\cos (p+\half k)_{\nu} \nnn
&&\mbox{}\frac{1}{16}
\Tr \{ \Xi_5\Gm_5 \Xi_{\mu} S(p)\Xi_5\Gm_5 \Xi_{\mu} S(p+k) \} \nnn
&=& \half y_R^2 N_D |h(k)|^2
e^{i\half (k_{\mu} - k_{\nu})} \sum_p
\frac{\cos (p+\half k)_{\mu}\cos (p+\half k)_{\nu}}{D(p) D(p+k)}
\left\{ -\dl_{\mu\nu}[\sin p \cdot \sin (p+k) \phantom{\quart} \right.\nnn
&& \left. \mbox{}+ \quart m_F^2 \cos p \cdot
\cos (p+k)] + \l_{\mu\nu} m_F^2 [\cos p_{\mu} \cos (p+k)_{\nu} +
\mu \leftrightarrow \nu] \right\}\;.
\eea
For $k=0$ this reduces to
\be
\Sg_{\mu\nu}^{(c)}(0) = \half y_R^2 N_D
\left[ -\dl_{\mu\nu} \sum_p \frac{c_1^2 }{D} +
l_{\mu\nu} 2 m_F^2 \sum_p \frac{c_{\mu}^2 c_{\nu}^2}{D^2} \right],
\label{SIGB0}
\ee
where $D$ was defined in (\ref{DDEF}) and we use the notation
\be
c_{\mu} = \cos p_{\mu},\;\;\;s_{\mu} = \sin p_{\mu},\;\;\;
s^2 = \sum_{\mu} \sin^2 p_{\mu}\; .
\ee

The total fermion contribution to the scalar self energy equals
$\Sg_{\mu\nu}(k)=\Sg_{\mu\nu}^{(b)}(k) + \Sg_{\mu\nu}^{(c)}(k)$.
In an O(4) symmetric model the transverse
parts of the zero momentum
$\Sg_{\mu\nu}^{(b)}$ and $\Sg_{\mu\nu}^{(c)}$
in (\ref{TADPO}) and (\ref{SIGB0}) would cancel
such that $m_{\pi}$ remains zero. Using $\dl_{\mu\nu} = l_{\mu\nu}
+ t_{\mu\nu}$ we see such a cancellation in the first term of
(\ref{SIGB0}). However, despite its factor $l_{\mu\nu}$, the
second term also contains a transverse part, as follows from
\be
l_{\mu\nu} \sum_p \frac{c_{\mu}^2 c_{\nu}^2}{D^2} =
l_{\mu\nu}\sum_p \frac{c_1^2 c_2^2 }{D^2} +
\dl_{\mu\nu}\quart\sum_p \frac{c_1^4 - c_1^2 c_2^2}{D^2}.
\ee
So we find
\bea
\Sg_{\mu\nu}(0)&=&l_{\mu\nu} \left[ \half y_R^2 N_D \sum_p
(2\frac{c_1^2 }{D} + 2 m_F^2 \frac{c_1^2 c_2^2 }{D^2})
+ 2\vep_R v_R^2 \right] + t_{\mu\nu} 2\vep_R v_R^2\; ,\\
\vep_R&=& y_R^4 \frac{N_D}{8}\sum_p \frac{c_1^4 - c_1^2 c_2^2}{D^2}
=f_{\vep} N_D y_R^4\;, \label{EPDEF}
\eea
leading to a pion mass $\propto \vep_R$.
The values of $f_{\vep}$ range from 0.0054 to 0.0043
for $m_F$ ranging from 0 to 0.5.   \\

In ref.~\cite{BoSm92} we followed a slightly different strategy and
computed the coefficient $\vep_R$ of the
term $\vep_R\Sm\vr_{R\mu}^4$ in the effective
potential (\ref{EFFERPOT}). There
we found the same coefficient $\vep_R$ as computed here from the
two point function. \\

The continuum limit of $\Sg_{\mu\nu}(k)$ can be calculated
by separating the integration region ($\sum_p \ra \int_{-\pi/2}^{\pi/2}
d^4 p/\pi^4$) into a small ball around the origin and the
outer region (see e.g. \cite{GoSm84}), which leads to the
form
\bea
\Sg_{\mu\nu}(k) &=& y_R^2 N_D \left\{ l_{\mu\nu}(c_{-2} + c_0 m_F^2
-\frac{3m_F^2}{2\pi^2}\ln m_F^2) + \dl_{\mu\nu} k^2
(\ta -\frac{1}{4\pi^2}\ln m_F^2) \right.   \nnn
&&\left. \mbox{} -\frac{1}{2\pi^2}\int_0^1 dx \left[\dl_{\mu\nu} (m_F^2 +
3x(1-x) k^2) + l_{\mu\nu} 2m_F^2\right]\ln
\left[1+x(1-x)\frac{k^2}{m_F^2}\right] \right\}\nnn
&&\mbox{}
+ 2\vep_R v_R^2 \dl_{\mu\nu} + \dl_R \dl_{\mu\nu} k_{\mu}^2\; .
\label{SIGCONT}
\eea
Here $c_{-2}$, $c_0$ and $\ta$ are numerical coefficients
which we have not calculated explicitly.
The term $\dl_R \dl_{\mu\nu} k_{\mu}^2$ corresponds to the second
O(4) symmetry breaking term of dimension four in eq.~(\eq{C2}).
%
%
The coefficient $\dl_R$ can be
expressed as
\bea
\dl_R &=& \half y_R^2 N_D \sum_p \left [\frac{ \frac{1}{4}(c_1^2-s_1^2) }
{D} - a \frac{ \frac{3}{2} c_1^2 s_1^2-\frac{1}{2} c_1^2 s_2^2 -c_1^4
+ c_1^2 c_2^2 } {D^2} -2a^2 \frac{c_1^4 s_1^2-c_1^2 c_2^2 s_2^2}{D^3}
\right. \nonumber \\
&+&\frac{m_F^2}{4} \left\{ \frac{ \frac{5}{2} c_1^2 s_1^2 -c_1^2 s_2^2
-\frac{3}{2} c_1^2 (c_1^2-c_2^2) }{D^2}
+ a \frac{2 c_1^4 (-c_1^2+3 c_2^2 +5 s_1^2-s_2^2)-4 c_1^2 c_2^2
(c_3^2 +3 s_1^2 -s_3^2 )}{D^3}  \right. \nonumber \\
& &  \left. \left. +a^2 \frac{8
c_1^4 (c_1^2 s_1^2 - c_2^2 s_2^2)-16c_1^2 c_2^2
(c_1^2 s_1^2 - c_3^2 s_3^2 )}{D^4} \right\} \right]
= f_{\dl} N_D y_R^2. \label{FDELTA}
\eea
with $a = 1-m_F^2/4$
(The expression (\ref{FDELTA}) differs from that in ref.
\cite{BoSm92} which
was incomplete). The values of $f_{\dl}$ range from 0.0026 to 0.0017
for $m_F$ ranging from 0 to 0.5.   \\

The quadratic and logarithmic divergencies of the continuum limit
correspond to the $c_{-2}$ and $\ln m_F^2$ terms respectively.
Note in particular that $\dl_R$ and $\vep_R$ are finite (the
continuum limit of $\vep_R$ and $\dl_R$ corresponds to letting $m_F\ra 0$ in
(\ref{EPDEF}) and (\ref{FDELTA})).
This is because the integrand near the origin in
momentum space is covariant, as in the continuum.
At the two loop level $\vep_R$ and $\dl_R$ probably acquire
logarithmic divergencies with corresponding renormalization group
$\beta$-functions. Presumably this means that $\vep_R$ and $\dl_R$
will be further suppressed by triviality. For example, in pure O(N)
models $\beta$-functions for asymmetric couplings have been studied
previously \cc{ZJetal}, and from these results we found it
amusing to discover that the ratio $\vep_R/\lm_R$ in the
asymmetric O(4) model vanishes logarithmically slowly in the
scaling limit.\\

We now perform mass and wave function renormalization. A change
in the parameter $m_R^2$ in (\ref{SR}) allows for a
renormalization of $c_{-2}$ like terms and a change of scale of
the scalar field allows for a renormalization of the $\ta$-like
terms. We may then write for the renormalized propagator
\bea
G_{R\mu\nu}^{-1}(k) &=& 2(\lm_R^{\prime} + \vep_R) v_R^2 l_{\mu\nu} +
2\vep_R v_R^2 t_{\mu\nu} + k^2\dl_{\mu\nu} +
\dl_R\dl_{\mu\nu} k_{\mu}^2 +
\Sg_{\mu\nu}^{(sub)} (k)\;,\label{GIREN}\\
\Sg_{\mu\nu}^{(sub)} (k)&=&\Sg_{\mu\nu}(k)-\Sg_{\mu\nu}(0)
- k_{\rh}^2 \left[ \frac{\partial}{\partial k_{\rh}^2}
\Sg_{\mu\nu} \right]_{k=0} \; . \label{SIGSUB}
\eea
Here we {\em defined} $\lm_R^{\prime}$ by its appearance in (\ref{GIREN}).
This definition is based on the effective tree level action
\be
S_{eff} \approx -\int d^4 x \left[  \frac{1}{2} \dm \vr_{R\nu}\dm \vr_{R\nu} +
\frac{m_R^2}{2} \vr_{R\mu}\vr_{R\mu} +
\frac{\lm_R^{\prime}}{4}(\vr_{R\mu}\vr_{R\mu})^2 +
\sum_{\mu}[\frac{\dl_R}{2} (\dm \vr_{R\mu})^2 + \vep_R \vr_{R\mu}^4] \right] \;
,
\ee
which leads to (\ref{GIREN}) with $\Sg^{(sub)}\ra 0$.\\

To determine the scalar propagator from (\ref{GIREN}) we
use the $\sg,\pi$ basis,
\bea
G_R^{\al\bt}(k)&=& e_{\mu}^{\al}e_{\nu}^{\bt}
G_{R\mu\nu}(k),\;\;\; \Sg_{(sub)}^{\al\bt}(k) =
e_{\mu}^{\al}e_{\nu}^{\bt}\Sg_{\mu\nu}^{(sub)} (k) \lb{GSYM1} \\
G_{\sg R} (k) &\equiv& l_{\mu \nu} G_{\mu \nu}(k) =
\left[ 2(\lm_R^{\prime} + \vep_R) v_R^2 + (1+\frac{\dl_R}{4})k^2 +
      \Sigma_{(sub)}^{44}(k) + O(\dl_R^2) \right]^{-1}, \lb{GSYM2} \\
G_{\pi R} (k) &\equiv& \third t_{\mu \nu} G_{\mu \nu}= \third\sum_{a=1}^3
\left[ 2\vep_R v_R^2 + (1+\frac{\dl_R}{4})k^2 +
      \Sigma_{(sub)}^{aa}(k) + O(\dl_R^2) \right]^{-1} \;.
\lb{GSYM3}
\eea
In the $\sg$, $\pi$ basis the transverse and longitudinal projectors are
of course diagonal, but the $\dl_R$ term has an off-diagonal
piece. To establish (\ref{GSYM2}) and (\ref{GSYM3})
for the scalar propagators it is
convenient to use the basis vectors  $e^1=\half(1,-1,-1,1)$,
$e^2=\half(-1,1,-1,1)$ and $e^3=\half(-1,-1,1,1)$ for the modes
orthogonal to $e^4=\half(1,1,1,1)$. With this choice the
$\sg,\pi$ diagonal part of the $\dl_R$ term takes an O(4)
invariant form, $\dl_R\dl^{\al\bt}k^2/4$, and
gives an additional wave-function renormalization by a factor
$1+\dl_R/4$. The off-diagonal terms
$ \dl_R\Sm (e^{\al}_{\mu}e^{\bt}_{\mu} -\dl^{\al\bt}/4)k_{\mu}^2$
lead to corrections
of order $\dl_R^2$ in $G_{\sg,\pi}$ which may be neglected.
\\

{}From eq. (\ref{GSYM2}) and (\ref{GSYM3}) we read off the following one
fermion loop
estimate for the Higgs and Goldstone masses,
\be
m_{\sg}^2=\frac{2(\lm_R^{\prime} +\vep_R) v_R^2}
                  {1+\dl_R/4}\;,\;\;
m_{\pi}^2=\frac{2 \vep_R v_R^2}
                  {1+\dl_R/4} \;.  \lb{M}
\ee
In ref.~\cite{BoSm92} we found that this result for the Goldstone mass
was in reasonable agreement with the numerical data, which will
be supported by our results in sect. 4.
This motivates us to take also the above result for the Higgs mass
seriously and use it for an improved definition of the quartic
coupling which is corrected for the O(4) symmetry breaking,
\be
 \lm_R^{\prime} =
\frac{m_{\sg}^2}{2v_R^2}(1-\frac{m_{\pi}^2}{m_{\sg}^2})
(1+\frac{\dl_R}{4}) \;. \lb{LMCORR}
\ee
We shall compare this with the usual definition
\be
\lm_R=\frac{m_{\sg}^2}{2v_R^2}
\ee
in our numerical work to be discussed below.
After checking that the corrections are small,
we may neglect them.\\

To this one loop order the coupling $y_R$ may be interpreted to
include also the fermion feedback on the scalars. In the spirit
of renormalized perturbation theory, (\ref{GSYM2}) and (\ref{GSYM3})
are expected to
be valid whenever the renormalized couplings are sufficiently
small, even when the bare couplings are large. \\

It can be seen from (\ref{SIGCONT}) and (\ref{SIGSUB}) that
$\Sg^{(sub)}$ becomes important for $k^2\ageq 4m_F^2$.
We shall use expressions (\ref{GSYM2}) and (\ref{GSYM3}) in analyzing our
scalar propagator data, using for $\Sg_{(sub)}$ its full lattice
form and $k^2_{\rho} \ra \khat^2_{\rho}$ as in (\ref{SCALPROP}).
\section{Finite volume results for the fermion and scalar propagators}
In this section we turn to the numerical simulations.
After reporting on some technical details, we
discuss our methods for analyzing the fermion and scalar propagators
and we present the finite volume results.

We have measured the propagators
in three different regions of the bare $(\kp,y)$ parameter space,
$\kp=0.29$-$0.31$, $y=0.7$ (I), $\kp=0$, $y=3.6$-$4.8$ (II)
and $\kp=-0.25$, $y=5.6$-$5.8$ (III). These  regions with the points at
which we have performed the simulations have been indicated in the phase
diagram of fig.~\ref{Fphased}.
The region (I) was included to get a crude idea of
the $y_R$ dependence of $\lm_R$
for intermediate values of the Yukawa coupling.
The regions (II) and (III) were chosen to study the model at
large values of the bare Yukawa coupling. Region (II) would
correspond to $y_0=\infty$
in the continuum parametrization of the action given in eq.~(\ref{S0F}).

We have restricted our calculations to the FM phase,
in particular, we did not increase $y$ and lower $\kp$ further beyond
region (III), since from comparing the results in region (II) and (III)
we observed that the renormalized couplings
$\lm_R$ and $y_R$ remain almost constant
when lowering $\kp$ beyond $\kp=0$, while
keeping $a v_R$ roughly fixed. We expect this trend to continue
also beyond the multicritical point A.
The bosonic particle
spectrum in the FI phase, however, differs from that in the FM phase.
It contains
in addition to the usual Higgs and Goldstone particles
so-called `staggered' Higgs and Goldstone particles which
are associated with the antiferromagnetic ordering in that phase \cc{BoDe91}.

We carried out the simulation on volumes
of size $V=L^3 \times T$ where
the spatial extent ranged from $L=6$ to $16$
and the time extent was always
kept fixed at $T=24$.
The trajectory length $\tau$ in the HMCA was set equal to one
and we have tuned
the step size $\dl \tau$ such that the acceptance rate fell into a range
between $70$ and $80$\%.
In general $\dl \tau$ had to be reduced slightly when
the value of $\kp$ was lowered
and the value of $y$ was raised. We used for the fermions
periodic boundary conditions
in spatial directions and antiperiodic boundary conditions
in the time direction. The scalar fields had periodic
boundary conditions in all directions.
Depending on the autocorrelation times for the scalar
field expectation value we have generated 5,000
to 20,000 trajectories, which resulted in reasonably
small statistical errors for the various observables.
We could afford large lattices and high
statistics because the staggered fermion matrix in eq.~(\eq{SCHI})
is relatively small and the conjugate gradient inversions
were found to converge excellently.
%
%
\begin{figure}
\centerline{
\fpsxsize=10.cm
\fpsbox{Fpropf.ps}
}
\caption{ \noindent {\em The quantity
\protect{$\sin p_4/S^{\Gm}_4(p_4)$} as a function of
\protect{$\sin^2 p_4$}
for $\kp=0$ and several values of $y$.
The lattice size is $12^3 \times 24$. The straight lines were obtained
by fitting the Ansatz ({\protect \eq{FAN}}) to the numerical results.
}}
\label{Fpropf}
\end{figure}
\subsection{Fermion propagator}
The fermion propagator in the momentum space representation is
defined by the expression
\be
  S_{AB}(p) = \left\lag  \frac{1}{V}
   \sum_{x,y} e^{i(p+\pi_A)x} M_{xy}^{-1} e^{-i(p+\pi_B)y}
   \right\rag \;, \lb{SAB}
\ee
where $M_{xy}$ is the fermion matrix defined in (\ref{SCHI}).
Assuming that loop effects are small, we can use the
free fermion result in eq.~(\ref{SABREN}) to parametrize $S_{AB}$.
This suggests to measure the projections
\be
S^{\Gm}_{\mu}= i \textstyle{
\frac{1}{16}}\Tr \{ \Gm_{\mu}S \}            \;,\;\;\;
S^{\Xi}_{\mu} = \textstyle{\frac{1}{16}}\Tr \{\Gm_5\Xi_5\Xi_{\mu}S \}  \;.
\ee
On a finite volume, the scalar field  will tunnel from one minimum
of the effective potential to another, where the average field
value $\Sx \vr_{\mu x}/V$ is proportional to
$(\pm 1,\pm 1,\pm 1,\pm 1)$ in these minima.
The propagator component $S^{\Gm}_{\mu}$ is the same in all minima, but
$S^{\Xi}_{\mu}$ is proportional to $\Sx \vr_{\mu x}/V$
and would vanish in  a finite volume.\\

In practise we have measured $S^{\Gm}_4(p)$
by averaging over rotated scalar field configurations.
These global O(4) rotations were chosen such that
$\Sx \vr_{\mu x}/V \propto \dl_{\mu 4}$ on all configurations that
went into the averaging. The fermion momenta were chosen along
the 4-direction:
\be
p_4= \left(n-\frac{1}{2}\right) \frac{2\pi}{T}\;,\;\;\;\;
n=-\frac{T}{4}+1,\ldots, \frac{T}{4}
\ee
and $p_j=0$.
Typical results are shown in fig.~\ref{Fpropf}, where we
have plotted $\sin p_4/S^{\Gm}_4(p_4)$ for several values of $y$ and $\kp=0$
as a function of $\sin^2 p_4$.
The linear behavior which is characteristic for weakly interacting fermions
is seen to hold remarkably well over the full range of momenta. \\

In order to extract the fermion mass from these data we
adopt the free fermion parametrization,
\be
 \frac {   \sin p_4   }{   S^{\Gm}_4(p_4) }
 \approx \frac {(1-m_F^2)\sin^2 p_4 + m_F^2 }{Z_F} \;,
\lb{FAN}
\ee
and fitted the parameters $m_F$ and $Z_F$.
This formula follows from (\ref{SABREN}) with the modification
that the vector $e^4_{\mu}$ in the mass part, which corresponds to
the direction of symmetry breaking, now is the unit vector in the
4-direction. \\

The results of the fits are collected in table~\ref{Tpropf}.
The fermion mass found from $S^{\Xi}_4$ with the
appropriate free fermion fit gave the same results within 10\%.
This small discrepancy decreased on larger
lattices, presumably because $\lag \varphi_{\mu} \rag$ gets
frozen in one of the minima.
\begin{table}
\begin{center}
\begin{tabular}{|r|l|l|r|l|l|l|l|} \hline
  &$\kappa$&$y$&$L$&$m_F$&$Z_F$&$y_R$&$\yt$ \\
                                                       \hline\hline
 (I)&$0.29$&$0.7$&$ 6$&$0.1844(1)$&$0.9768(6)$&$0.69(2)$&$0.72(3)$ \\
  &      &     &$ 8$&$0.1496(7)$&$0.980(2) $&$0.79(4)$&$0.72(8)$ \\
  &      &     &$10$&$0.129(1) $&$0.981(4) $&$0.90(7)$&$0.6(2) $ \\
 \hline
  &$0.30$&     &$ 6$&$0.230(1) $&$0.976(2) $&$0.75(2)$&$0.77(5)$ \\
  &      &     &$ 8$&$0.2074(4)$&$0.9794(6)$&$0.82(2)$&$0.75(9)$ \\
  &      &     &$10$&$0.198(1) $&$0.9807(2)$&$0.85(3)$&$0.6(2) $ \\
  &      &     &$12$&$0.1940(4)$&$0.9809(2)$&$0.83(2)$&$0.9(2) $ \\
 \hline
  &$0.31$&     &$ 6$&$0.2672(3)$&$0.976(4) $&$0.77(2)$&$0.80(6)$ \\
  &      &     &$ 8$&$0.2523(5)$&$0.980(1) $&$0.81(2)$&$0.8(1) $ \\
  &      &     &$10$&$0.2473(6)$&$0.9813(4)$&$0.84(3)$&$0.6(4) $ \\
  &      &     &$12$&$0.2446(6)$&$0.9816(3)$&$0.84(2)$&$0.5(4) $ \\
 \hline
(II)&$0.00$&$3.6$&$12$&$0.2004(8)$&$0.847(1) $&$2.32(7)$&$1.89(3)$ \\
 \hline
  &       &$3.8$&$ 6$&$0.488(1)$&$0.800(4) $&$2.61(6)$&$2.37(2)$ \\
  &       &     &$ 8$&$0.394(2)$&$0.821(7) $&$2.51(4)$&$2.25(4)$ \\
  &       &     &$10$&$0.355(3)$&$0.8259(6)$&$2.48(4)$&$2.33(4)$ \\
  &       &     &$12$&$0.333(3)$&$0.826(3) $&$2.48(5)$&$2.34(4)$ \\
  &       &     &$16$&$0.318(2)$&$0.825(1) $&$2.50(5)$&$2.28(8)$ \\
 \hline
  &      &$4.0$&$ 6$&$0.5804(3)$&$0.777(8) $&$2.85(5)$&$2.50(9)$ \\
  &      &     &$ 8$&$0.509(1) $&$0.799(3) $&$2.75(2)$&$2.50(4)$ \\
  &      &     &$10$&$0.491(5) $&$0.805(6) $&$2.72(4)$&$2.52(7)$ \\
  &      &     &$12$&$0.482(2) $&$0.807(2) $&$2.68(5)$&$2.58(9)$ \\
 \hline
  &     &$4.2$&$ 6$&$0.674(5)  $&$0.75(1)  $&$2.95(6)$&$2.93(9)$ \\
  &     &     &$12$&$0.624(1)  $&$0.790(2) $&$2.91(5)$&$2.7(1) $ \\
 \hline
  &     &$4.8$&$12$&$0.9839(5) $&$0.7596(7)$&$3.36(5)$&$3.3(2) $ \\
 \hline
(III)&$-0.25$&$5.6$&$ 6$&$0.492(1)$&$0.775(5)$&$2.83(4)$&$2.42(3)$ \\
   &       &     &$ 8$&$0.400(1)$&$0.794(2)$&$2.65(4)$&$2.41(4)$ \\
   &       &     &$10$&$0.360(3)$&$0.799(2)$&$2.68(4)$&$2.30(5)$ \\
   &       &     &$12$&$0.338(4)$&$0.799(2)$&$2.61(4)$&$2.42(2)$ \\
 \hline
   &      &$5.8$&$ 6$&$0.5480(3)$&$0.758(4)$&$3.02(4)$&$2.51(5)$ \\
   &      &     &$ 8$&$0.475(2) $&$0.778(3)$&$2.85(4)$&$2.58(4)$ \\
   &      &     &$10$&$0.448(5) $&$0.785(3)$&$2.75(4)$&$2.63(5)$ \\
   &      &     &$12$&$0.446(1) $&$0.781(3)$&$2.84(3)$&$2.66(5)$ \\
\hline\hline
\end{tabular}
\caption{{ \em
Results for the fermionic observables $m_F$, $Z_F$, $y_R$ and $\yt$
in the  regions (I), (II) and (III) of the phase diagram
(c.f. fig.~{\protect \ref{Fphased}})
for several values of $L$ and $T=24$.
}}
\label{Tpropf}
\end{center}
\end{table}
\subsection{Scalar propagator}
We have measured the  following momentum space Green functions,
\be
G_{\sg}(p) = G_{44}(p),\;\;\;
G_{\pi}(p)=\frac{1}{3} \sum_{m=1}^3 G_{m m}(p) \;,    \lb{GSGPI}
\ee
with
\be
  G_{\mu\nu}(p) = \left\lag
  \frac{1}{V} \sum_{xy} \vr_{\mu x}\vr_{\nu y}e^{ip(x-y)}
\right\rag \;,
\ee
which is identical to (\eq{GGSS}) for non-zero momenta $p$.
The scalar fields  in $G_{\mu \nu}(p)$ have been rotated such that
$\Sx \vr_x/V\propto \dl_{\mu 4}$, as was done for the computation
of the fermion propagator.
For the definition of $G_{\sg , \pi}$ in (\eq{GSYM1})-(\eq{GSYM3}) this
is just a change of the basis such that $e_{\mu}^4 \ra \dl_{\mu 4}$,
$G_{\sg} =l_{\mu \nu} G_{\mu \nu} \ra G_{44}$,
$G_{\pi} =\third t_{\mu \nu} G_{\mu \nu} \ra
\third \sum_{m=1}^3  G_{m m} $, as in (\eq{GSGPI}). Note, however,
that our fitting formulas (\eq{GSYM2}) and (\eq{GSYM3}) should be
unchanged, i.~e. we still use in $\Sg_{(sub)}$ the $e_{\mu}^4$ of
eq.~(\eq{E4MU}) (The dynamical fermions in the Monte Carlo
process `experience' of course the preferred local minima (\eq{E4MU}) and
not $e_{\mu}^4=\dl_{\mu 4}$). \\

In fig.~\ref{Fprops} we show a representative example of the measured
inverse propagators $G_{\sg,\pi}(p)^{-1}$, as a function of the
bosonic lattice version of the momentum squared,
$\phat= 2 \Sm (1-\cos p_{\mu})$. To reduce statistical errors
we have computed the  propagator in the small momentum range for
all possible lattice momenta, and averaged over results with the same
value of $\phat$. \\

For sufficiently small momenta, $\phat \ll m_F^2$, $m_{\sg,\pi}$
one expects a linear $\phat$ dependence, $G_{\sg,\pi}^{-1} =
(\phat + m_{\sg,\pi}^2)/Z_{\sg,\pi}$ and a naive linear fit was
commonly applied to extract $m_{\sg,\pi}$ and $Z_{\sg,\pi}$
\cc{FrLi92,BoSm92,Search}.
However, the data in fig.~\ref{Fprops} show a significant curvature,
which may give a bias in an estimate of $m_{\sg,\pi}$ and $Z_{\sg,\pi}$,
obtained from a linear fit.
This bias can be large on small lattices, where even the smallest
momenta are too large to neglect the curvature.
Therefore we followed
the same strategy as in ref.~\cite{BoDe91}
and included the subtracted self-energy $\Sigma_{(sub)}(p)$
(cf.~eqs.~(\eq{GSYM2}) and (\eq{GSYM3}))
to parametrize the non-linear $\phat$ dependence.
Furthermore it was found that the bending
over of $G_{\sg,\pi}^{-1}$ for very large momenta,
$\phat \ageq 4$ (see fig.~\ref{FpropsB}), could
be reproduced by including the pole contribution coming from the
`staggered' scalar modes. Apart from the usual
Higgs and three Goldstone bosons the spectrum in the FM phase contains also
four `staggered' scalar modes which are associated with the
antiferromagnetic ordering in the FI phase. These modes
produce a pole in the diagonally opposite corner of the Brillouin zone,
at $16 -\phat = (m^{st}_{\Phi})^2$.
The mass $m^{st}_{\Phi}$ of these particles is usually large, but it
becomes small when approaching point A in the
phase diagram. See ref.~\cite{BoDe91} for a detailed discussion of
the staggered scalar modes.  \\

These considerations lead us to use the following parametrization for
the inverse scalar propagators in the FM phase,
\be
 G_{\sg,\pi}(p) \approx
Z_{\sg,\pi}/
\left(\phat + m_{\sg,\pi}^2 +\frac{\yt^2}{y_R^2}\Sigma_{(sub)}(p)\right)
       +  Z^{st}_{\Phi}/(16-\phat + (m^{st}_{\Phi})^2) \;.
        \lb{FITS}
\ee
This requires us to fit five parameters: the mass and wave-function
renormalization factors for the scalar particles ($\sg$ or $\pi$) and for
the staggered
bosons and the coefficient of the subtracted self-energy, $\yt$.
This last parameter measures the strength of the renormalized three point
Yukawa interaction and should be close to $y_R$ obtained from the
tree level relation with the fermion mass.
Notice that $\Sigma_{(sub)}/y_R^2$ contains no free parameters, since
we use the measured value of the fermion mass in the loop sum.
The subtracted self-energy $\Sigma_{(sub)}(p)$ is computed
numerically for the lattice sizes used in the simulations.
The coefficient of the term linear in $\phat$ is computed with
the finite difference approximation
$[\partial
\Sigma(k)/\partial k_{\rho}^2
]_{k=0} \approx (\Sigma(h_{4})-\Sigma(0))/{\hat{h}_4}^2$,
where  $h_{4}$ is the smallest non-zero lattice momentum  in the
time direction.\\
%
\begin{figure}
\centerline{
\fpsxsize=11.5cm
\fpsbox{Fsi038.ps}
}
\vspace*{-1.0cm}
\centerline{
\fpsxsize=11.5cm
\fpsbox{Fpi038.ps}
}
\caption{ \noindent {\em
The inverse scalar propagators $G_{\sg}^{-1}(p)$ (fig.~a)
and $G_{\pi}^{-1}(p)$ (fig.~b)
as a function of $\phat$ at the point $(\kp,y)=(0,3.8)$. The lattice
size is $12^3 24$. The circles
were obtained by fitting
the Ansatz ({\protect\eq{FITS}}) to
the Monte Carlo data which are represented by the
crosses.
}}
\label{Fprops}
\end{figure}
%
%
%
\begin{figure}
\centerline{
\fpsxsize=18.0cm
\fpsbox{Fprops.ps}
 }
\caption{ \noindent {\em The inverse scalar propagator
$G_{\pi}(p)$ is plotted for the full momentum range
as a function of $\phat$. The computation was performed at the point
$(\kp,y)=(-0.25,5.6)$ (region (III)) on a lattice of size $12^3 24$.
The points in the lower figure were obtained by fitting
the Ansatz ({\protect\eq{FITS}}) to the
Monte Carlo data which are displayed in the upper figure.
  }}
\label{FpropsB}
\end{figure}
%
%

If the renormalized Yukawa coupling is weak, one expects
that the one-loop expression (\ref{FITS}) gives a good description
of the non-linear $\phat$ dependence of the inverse propagator.
Conversely, an accurate description of the data over a wide
momentum range with the one-loop form (\ref{FITS}) indicates
that the Yukawa coupling must be weak.  As will be seen shortly,
this is the case in our model and including the self-energy
term leads to stable values for the mass and wave-function
renormalization factor even on small lattices.\\

%
%
%
%
\begin{table}
\begin{center}
\begin{tabular}{|r|l|l|r|l|l|l|l|l|l|} \hline
 &$\kp$&$y$&$L$&$v_R$&$m_{\pi}$&$m_{\sg}$&
$Z_{\pi}$&$Z_{\sg}$&$\rmv$ \\
                                                             \hline\hline
 (I) & $ 0.29$&$0.7$&$ 6$&$0.267(9)$&$0.09(4)$&$0.58(2)$&$0.76(5) $&$1.12(2)
$&$2.2(1) $   \\
   &        &     &$ 8$&$0.19(1) $&$0.17(5)$&$0.44(1)$&$1.0(1)  $&$1.13(1)
$&$2.4(2) $   \\
   &        &     &$10$&$0.14(1) $&$0.18(6)$&$0.38(2)$&$1.3(2)  $&$1.17(2)
$&$2.6(2) $   \\ \hline
   & $ 0.30$&     &$ 6$&$0.31(1) $&$0.13(3)$&$0.72(3)$&$0.96(6) $&$1.23(3)
$&$2.4(1) $    \\
   &        &     &$ 8$&$0.254(7)$&$0.14(3)$&$0.63(2)$&$1.19(6) $&$1.25(2)
$&$2.48(9)$   \\
   &        &     &$10$&$0.235(8)$&$0.17(4)$&$0.65(2)$&$1.32(9) $&$1.31(2)
$&$2.8(1) $    \\
   &        &     &$12$&$0.235(6)$&$0.13(4)$&$0.65(1)$&$1.28(7) $&$1.33(2)
$&$2.75(9)$   \\ \hline
   & $ 0.31$&     &$ 6$&$0.346(7)$&$0.13(3)$&$0.84(3)$&$1.04(4) $&$1.20(4)
$&$2.4(1) $    \\
   &        &     &$ 8$&$0.313(8)$&$0.10(4)$&$0.80(3)$&$1.21(6) $&$1.23(4)
$&$2.6(1) $    \\
   &        &     &$10$&$0.30(1) $&$0.14(4)$&$0.84(2)$&$1.34(9) $&$1.28(2)
$&$2.8(1) $    \\
   &        &     &$12$&$0.290(7)$&$0.14(4)$&$0.81(2)$&$1.38(7) $&$1.27(2)
$&$2.79(9)$   \\ \hline
(II) & $ 0.00$&$3.6$&$12$&$0.086(3)$&$0.15(2)$&$0.28(2)$&$0.26(1)
$&$0.289(3)$&$3.2(3) $   \\  \hline
   &        &$3.8$&$ 6$&$0.187(4)$&$0.15(1)$&$0.74(4)$&$0.34(1) $&$0.56(1)
$&$4.0(2) $               \\
   &        &     &$ 8$&$0.157(3)$&$0.16(2)$&$0.64(1)$&$0.35(1)
$&$0.500(4)$&$4.1(1) $    \\
   &        &     &$10$&$0.143(2)$&$0.11(2)$&$0.55(2)$&$0.35(1)
$&$0.464(5)$&$3.9(1) $    \\
   &        &     &$12$&$0.134(3)$&$0.11(3)$&$0.50(1)$&$0.35(1)
$&$0.441(3)$&$3.69(8)$   \\
   &        &     &$16$&$0.127(3)$&$0.17(3)$&$0.50(2)$&$0.36(1)
$&$0.410(7)$&$3.9(2) $    \\ \hline
   &        &$4.0$&$ 6$&$0.203(4)$&$0.18(2)$&$0.81(3)$&$0.40(1) $&$0.58(1)
$&$4.0(2) $               \\
   &        &     &$
8$&$0.185(2)$&$0.15(1)$&$0.72(1)$&$0.412(6)$&$0.555(6)$&$3.90(7)$
                                \\
   &        &
&$10$&$0.181(2)$&$0.12(2)$&$0.68(2)$&$0.414(9)$&$0.537(7)$&$3.8(1) $
                                 \\
   &        &     &$12$&$0.180(3)$&$0.11(4)$&$0.65(2)$&$0.41(1)
$&$0.528(9)$&$3.6(1) $    \\ \hline
   &        &$4.2$&$ 6$&$0.229(4)$&$0.14(2)$&$0.78(4)$&$0.41(1) $&$0.58(2)
$&$3.4(2) $               \\
   &        &     &$12$&$0.214(4)$&$0.18(2)$&$0.87(2)$&$0.47(2) $&$0.64(1)
$&$4.1(1) $               \\ \hline
   &        &$4.8$&$12$&$0.293(4)$&$0.29(2)$&$1.34(3)$&$0.54(2) $&$0.90(2)
$&$4.6(1) $               \\ \hline
(III) &
$-0.25$&$5.6$&$6$&$0.174(2)$&$0.17(1)$&$0.76(2)$&$0.180(4)$&$0.293(4)$&$4.4(1)
$                         \\
   &        &     &$
8$&$0.151(2)$&$0.11(2)$&$0.65(1)$&$0.175(4)$&$0.268(3)$&$4.34(9)$
                                \\
   &        &
&$10$&$0.135(2)$&$0.17(2)$&$0.59(1)$&$0.192(4)$&$0.251(2)$&$4.4(1) $
                                \\
   &        &
&$12$&$0.129(1)$&$0.13(1)$&$0.53(1)$&$0.185(2)$&$0.237(1)$&$4.13(6)$
                                \\ \hline
   &        &$5.8$&$
6$&$0.182(2)$&$0.18(1)$&$0.84(3)$&$0.209(4)$&$0.321(9)$&$4.6(2) $
                                \\
   &        &     &$
8$&$0.167(2)$&$0.16(2)$&$0.74(2)$&$0.207(4)$&$0.302(5)$&$4.4(1) $
                                \\
   &        &
&$10$&$0.163(2)$&$0.11(2)$&$0.69(2)$&$0.205(4)$&$0.288(6)$&$4.2(2) $
                                \\
   &        &
&$12$&$0.157(2)$&$0.16(1)$&$0.63(1)$&$0.213(3)$&$0.271(2)$&$4.00(7)$
                                \\
\hline \hline
\end{tabular}
\caption{ {\em
Results for the bosonic observables $v_R$, $m_{\sg,\pi}$, $Z_{\sg,\pi}$
and the ratio $m_{\sg}/v_R$ in the regions (I), (II) and (III) of the
phase diagram at several values of $L$ and $T=24$.
}}
\label{Tprops}
\end{center}
\end{table}
We applied this method of analysis to the scalar propagators
in all three regions of the phase diagram.
In region (I) we found larger
autocorrelations which urged us to increase the statistics to
20,000 trajectories.
Region (III) was not very different
from region (II). Since the AM phase is closer here, we found a stronger
effect of the staggered scalar modes on the scalar propagator for large
momenta, but this could still very well be fitted with the Ansatz
(\ref{FITS}).\\

We have fitted the inverse Goldstone propagators over the momentum range
$0 < \phat < 1.3$ and the $\sg$ particle propagator
over the interval $2(1-\cos (2 \pi / T)) < \phat < 1.7$.
In these intervals the effect of the staggered pole in eq.~(\eq{FITS})
is negligible and could be dropped. In fig.~\ref{Fprops} we show an
example for a such fit at a point which falls
into the region (II). The result of the fit is shown
by the open circles and it is seen that the agreement with the measured
data (represented by the crosses) is excellent. In fig.~\ref{FpropsB}b we show
the result of a fit over the full momentum range
at a point which belongs to the region (III), this time including the
staggered scalar pole. Again it is seen that the quality of the
fit is very good, even for large momenta. The resulting
values for $m_{\pi,\sg}$, $Z_{\pi,\sg}$ and $\yt$ are consistent with
those which were obtained from the fit over the
restricted momentum interval.
The fitted values for $m_{\sg,\pi}^2$ and $Z_{\sg,\pi}$
are collected in table \ref{Tprops}.
The fitted values for $\yt$ can be found
in table \ref{Tpropf}. They are  seen to be roughly equal to
the $y_R$ values,
which gives further support to the consistency of our analysis and
indicates that the renormalized Yukawa coupling is not very
strong. \\

By using the one fermion loop fit method we achieved a considerable
improvement of the scalar propagator analysis.
In ref.~\cite{BoSm92} we have used linear fits, but this was possible
only on the $12^3 24$ lattice since on smaller lattices
the systematic errors due to the
curvature at small momenta were unacceptably large. This problem
prohibited a study of the finite size effects and the extrapolation
to infinite volume. The linear fits on the $12^3 24$
lattice produced slightly larger values of $m_{\sg}$ and
smaller values of $Z_{\pi}$.
We emphasize that with the fitting method used
in this paper the systematic error for $m_{\sg}$ could be reduced by
almost an order of magnitude. \\

The renormalized couplings $\lm_R$ and $y_R$, listed in the tables
\ref{Tpropf} and \ref{Tprops}, were obtained from the tree
level relations (\eq{TRHIGGS}) and (\eq{YR}).
For the computation of these couplings we have to determine
the renormalized scalar field expectation value
\be
v_R=\frac{v}{\sqrt{ Z_{\pi} }} \;,  \lb{VR}
\ee
where $v$ is the unrenormalized scalar field expectation value.
As we mentioned in sect.~4.1, on a finite lattice the system is observed to
tunnel from one minimum to another and after averaging over many configurations
one would find zero for the scalar field expectation value. To compensate for
this drift it has proven to be useful to rotate the magnetization vector
for each configuration first into a certain direction, before taking
the assemble
average \cc{ROT}. On a finite lattice $v$ can then be defined by the relation
\be
v= \lag |\frac{1}{V} \sum_{x} \varphi_{\mu x}| \rag   \;. \label{DEFv}
\ee
As in the pure O(4) model we use $Z_{\pi}$ for the renormalization
of the scalar field expectation value in
eq.~(\eq{VR}), rather than $Z_{\sg}$, because the Higgs
particle is unstable. \\

One should be aware that the fitted value of
$Z_{\pi}$ contains the factor $(1+\dl_R/4)^{-1}$, cf.~(\ref{GSYM3}), which
implies that  $v_R^{\prime}=v_R (1+\dl_R/4)^{-1/2}$
should be used for the calculation of corrected couplings
$\lm_R^{\prime}$ and $y_R^{\prime}$ from
eqs.~(\eq{YR}) and (\eq{LMCORR}),
\bea
y_R^{\prime}&=&y_R (1+\frac{\dl_R}{4})^{1/2} \lb{YRC}\\
\lm_R^{\prime} &=&
\frac{m_{\sg}^2}{2v_R^2}(1-\frac{m_{\pi}^2}{m_{\sg}^2})
(1+\frac{\dl_R}{4})^2  \lb{LMRC}\;.
\eea
The ratios $R_y=(y_R-y_R^{\prime})/y_R^{\prime}$ and
$R_{\lm}=(\sqrt{2\lm_R}-\sqrt{2\lm_R^{\prime}})/ \sqrt{2\lm_R^{\prime}}$
give a measure for the O(4) symmetry breaking effects. These ratios were
computed numerically from eq.~(\eq{YRC}) and (\eq{LMRC}), where
the quantities $\ep_R$ and $\dl_R$ were determined
by means of eqs.~(\ref{EPDEF}) and (\ref{FDELTA}),
with the measured values of the fermion mass and $y_R$ supplied.
For the data listed in tables \ref{Tpropf} and \ref{Tprops} we find
\be
|R_y| <5\%  \;,\;\;\;\; |R_{\lm}| < 6\% \;, \lb{RRR}
\ee
in a parameter region with $m_F<0.5$ and $m_{\sg} <0.7$.
In general the numerical values of $|R_y|$ and $|R_{\lm}|$ decrease
with increasing lattice volume (see also fig.~\ref{Ffscorr}). \\

The accurate description of the momentum dependence of the
inverse scalar propagator over a wide momentum range,
using one fermion loop renormalized
perturbation theory indicates that  $y_R$ is in the perturbative
range and it suggests that also our computation of the O(4)
symmetry breaking effects should be reliable. This was found to be
the case in ref.~\cite{BoSm92}, where we compared the measured
values of the Goldstone mass with the one-loop result $m_{\pi}^2 \propto \ep_R$
of eq.~(\ref{M}). In fig.~\ref{Fmpi} we repeat this comparison using
our high statistics data. Again it is seen that the one-loop result gives a
reasonable description of the data. The increase of $m_{\pi}$
for $y$ values approaching the phase transition around $y\approx 3.6$ is due to
finite size rounding which is not contained in the one-loop formula.
Motivated by this result we shall assume that also
the one-loop corrections appearing in (\ref{YRC}) and
(\ref{LMRC}) give a reasonable account of the
remaining symmetry breaking effects.
%
%
\begin{figure}
\centerline{
\fpsxsize=10.0cm
\fpsbox{Fmpi.ps}
}
\caption{ \noindent {\em
The Goldstone mass as a function of $y$ for $\kp=0$. The lattice size
is $12^3 24$.
The measured values of $m_{\pi}$ are represented by the squares
and the one-loop result for $m_{\pi}$
obtained from eq.~({\protect\ref{M}})
by the diamonds.
}}
\label{Fmpi}
\end{figure}
\section{Infinite volume results for the renormalized couplings}
{}From the finite volume results for the fermion and Higgs mass we
want to obtain estimates of the renormalized couplings $y_R$ and
$\lm_R$ in the infinite volume. Therefore we need to extrapolate the
results of the previous section to the
infinite volume. In subsect.~5.1
we shall discuss how these extrapolations have been done.
The resulting infinite volume estimates
for the renormalized couplings are discussed in subsect.~5.2.
%
%
\begin{figure}
\centerline{
\fpsxsize=22.0cm
\fpsbox{Ffssup07.ps}
}
\vspace*{0.3cm}
\caption{ \noindent {\em The quantities $v_R$, $m_F$ and $m_{\sg}$
as a function of $1/L^2$ for the
three different coupling points of the region (I).
The extent of the
lattice in the time direction was kept fixed at $T=24$.
The dashed lines were obtained by a linear fit to all the data at different
lattices and the symbols drawn at
$1/L^2=0$ are the result of the infinite volume extrapolation. }}
\label{Ffssup07}
\end{figure}
%
%
%
\begin{figure}
\centerline{
\fpsxsize=22.0cm
\fpsbox{Ffssup.ps}
}
\vspace*{0.3cm}
\caption{ \noindent {\em
The same as fig.~8, but now for the regions (II) and (III).
}}
\label{Ffssup}
\end{figure}
\subsection{$1/L^2$ extrapolation to infinite volume}
If the spectrum
contains massless Goldstone bosons at infinite volume, this gives
rise to finite size effects in finite volume quantities which
vanish $\propto 1/L^2$. In ref. \cite{BoDe91} this was found to hold
for the scalar field expectation value, the Higgs mass and also
for the fermion mass. In our model the Goldstone particles
have a mass in an infinite volume due to the O(4)
symmetry breaking.
On a finite volume this is expected to give rise to deviations from the
linear $1/L^2$ dependence only when the lattice extent $L$
increases beyond the Goldstone correlation length, $L > O(1/m_{\pi})$.
Small deviations
imply that the symmetry breaking effects are small.
On the other hand, on the smaller lattices one expects
additional non-leading finite-size effects.
A pragmatic way to proceed is to apply the linear $1/L^2$
extrapolation
as long as no significant deviations are observed in the numerical data. \\

In fig.~\ref{Ffssup07} we plotted
the observables $v_R$, $m_{\sg}$ and $m_F$ obtained at the three different
points of region (I) as a function of $1/L^2$.
Fig.~\ref{Ffssup} shows the $1/L^2$ dependence of the same quantities
for the points in the regions (II) and (III).
The dashed lines are linear fits to
all data points. It is seen that the linear $1/L^2$
behavior holds satisfactory well within the statistical
error bars in all three regions of the phase diagram.
In region (I), where we expect the symmetry breaking effects
to be even smaller than in the other regions, the agreement
with a linear $1/L^2$ dependence is indeed best.
In regions (II) and (III), only the $m_F$ values at the points
which are most distant from the phase transition show a slight
indication of the expected systematic deviation for large volumes.
However, when comparing the results
obtained in regions (II) and (III) to the ones of region (I),
we conclude that this effect is rather small
and we interpret this observation as further evidence that
the effect of O(4) symmetry breaking in our model is small. \\

%
\begin{figure}
\centerline{
\fpsxsize=14.0cm
\fpsbox{Ffscorr.ps}
}
\caption{ \noindent {\em The
ratios $\protect{\sqrt{2 \lm_R}=m_{\sg}/v_R}$ (open symbols)
and $\protect{\sqrt{2 \lambda_R^{\prime}}}$ (filled symbols)
as a function of $1/L^2$ for $(\kp,y)=(0,4.0)$ (diamonds) and
$(-0.25,5.8)$ (squares).
The extent in the time direction was kept fixed at $T=24$.
The dashed lines (for the open symbols)
and the dotted lines (for the filled symbols)
were obtained by linear fits to all data at different
lattices and the symbols drawn at $1/L^2=0$ are the results of the
infinite volume extrapolation.
}}
\label{Ffscorr}
\end{figure}
Also the couplings themselves show a $1/L^2$ behavior as
can be seen in fig.~\ref{Ffscorr}, where we have displayed
as an example the ratio $\sqrt{2\lm_R}=m_{\sg}/v_R$ (open symbols)
as a function of $1/L^2$ for two  points
in region~(II) and~(III) in
the phase diagram. We showed in the previous
paragraph that both $m_{\sg}$ and $v_R$
are proportional to $1/L^2$, $v_R \simeq c_1 + c_2 /L^2$,
$m_{\sg} \simeq d_1 + d_2 /L^2$,  with relatively
small coefficients $c_2$ and $d_2$. From
an expansion in powers of $1/L^2$ we expect also
the ratio $m_{\sg}/v_R$ to obey an approximate $1/L^2$ behavior.
In addition we have also included in the graph
the results for the corrected ratio $\sqrt{2 \lambda_R^{\prime}}$
(filled symbols) which were
obtained by means of eq.~(\eq{LMRC}). Also here we expect, based
on an expansion of the right-hand side of eq.~(\eq{LMRC}) in powers of $1/L^2$,
an approximate linear $1/L^2$ dependence which is also seen
in the numerical results. Similar results hold also for the coupling
$y_R^{\prime}$ which we defined in eq.~(\eq{YRC}).
In general we find that the linear $1/L^2$
dependence is better fulfilled for the corrected ratios than
for the uncorrected ones, which gives some evidence that
the symmetry breaking effects in the corrected ratios
are small. It is amusing that in all the cases the
infinite volume extrapolations of the corrected and uncorrected ratios,
shown by the dotted and dashed lines in fig.~\ref{Ffscorr},
respectively, lead within the error bars to the
same values, though the numerical
values  on the smaller volumes deviate by two or sometimes three
standard deviations. \\

In order to estimate errors
we have performed extrapolations in various ways, e.g. we have used
subsets of the lattices
or we have either extrapolated the finite volume results
of the couplings or computed their infinite volume estimates
from the extrapolated infinite volume values of
$m_F$, $m_{\sg}$ and $v_R$. The resulting variations are
included in the errors quoted for
the infinite volume results.
In table \ref{Tinfres} we give the infinite volume results
for the renormalized couplings.
An estimation of
the symmetry breaking corrections as discussed in sect.~4
shows that these effects are small, less than 6\% in a parameter range
with $m_F <0.5$ and $m_{\sg}<0.7$.

%
\begin{table}
\begin{center}
\begin{tabular}{|r|l|l|l|l|l|l|} \hline
   &$  \kp$&$y  $&$m_{\sg} $&$v_R      $&$y_R     $&$\rmv   $ \\ \hline\hline
  (I)&$ 0.29$&$0.7$&$0.259(8)$&$0.08(2)  $&$1.00(9) $&$2.8(5) $ \\
   &$ 0.30$&     &$0.62(2) $&$0.21(2)  $&$0.87(3) $&$2.9(2) $ \\
   &$ 0.31$&     &$0.81(2) $&$0.271(2) $&$0.866(8)$&$2.96(8)$ \\ \hline
 (II)&$ 0.00$&$3.8$&$0.41(4) $&$0.1173(4)$&$2.45(4) $&$3.7(3) $ \\
   &       &$4.0$&$0.601(7)$&$0.170(5) $&$2.63(1) $&$3.6(2) $ \\ \hline
(III)&$-0.25$&$5.6$&$0.46(4) $&$0.114(2) $&$2.56(5) $&$4.1(3) $ \\
   &       &$5.8$&$0.56(4) $&$0.150(2) $&$2.7(1)  $&$3.8(3) $ \\ \hline\hline
\end{tabular}
\caption{{ \em
Infinite volume results for $m_{\sg}$, $v_R$, $y_R$ and $m_{\sg}/v_R$
for the various points in the regions (I), (II) and (III).
}}
\label{Tinfres}
\end{center}
\end{table}
\subsection{Renormalized couplings}
A comprehensive summary of our infinite volume results for the
renormalized couplings is
given in fig.~\ref{FLindn} where we have plotted
the ratio $m_{\sigma}/v_R=\sqrt{2\lm_R}$
as a function of $m_F/v_R=y_R$
for the various points of the regions
(I), (II) and (III). The symbols were chosen
such that they match with those in fig.~\ref{Fphased}
so that the reader can easily find out at which bare couplings
in the phase diagram the various points have been obtained.
All points have roughly the same value of the cut-off in units
of the scalar field expectation value, $a v_R \approx 0.15$-$0.25$.
The unsystematic scattering (as a function of the cut-off) of the points
in the region around $y_R=2.5$, $\sqrt{2 \lm_R}= 4$ is
still within one standard deviation and is presumably due
to uncertainties in the infinite volume extrapolation.
The figure shows that the data points obtained
in the regions (II) and (III) fall almost
on top of each other. The renormalized couplings appear to
saturate when $\kp \searrow 0$ and not to increase further
when $\kp$ is lowered beyond zero,
always keeping $av_R$ roughly fixed. \\

We furthermore
compare in fig.~\ref{FLindn} the
numerical results for $\sqrt{2 \lm_R}$ and
$y_R$ with the tree level unitarity bounds marked by the horizontal
and vertical arrows below and beside the axis.
The tree level unitarity bound
for $\sqrt{2 \lm_R}$ was taken from ref.~\cc{LuWe}. The bound on $y_R$
for a fermion-Higgs model with $N_D$ doublets is given by
\be
y_R \aleq \sqrt{4\pi/N_D}  \;,\lb{TUB}
\ee
i.~e. $y_R \aleq 2.5$ for $N_D=2$. The points
obtained in the regions (II) and
(III) are very close
to the tree level unitarity bounds which indicates that the
couplings are not very strong. This is also consistent with our
observation that the numerical results
for the scalar propagators could well be
described in the full momentum range by the
one fermion loop expressions. \\

For comparison we have displayed in
fig.~\ref{FLindn} also the $y_R$ dependence
of $\sqrt{2 \lm_R}$ using the assumption that the one-loop $\bt$-function
is a good approximation to the full $\bt$-function.
The one-loop $\bt$-functions
for the Yukawa and quartic self-coupling in a fermion-Higgs model with
$N_D$ doublets are given by the expressions:
\bea
\bt_y(\yba,\lmb) &\equiv& \frac{d \yba(t)}{dt} = \frac{N_D}{4\pi^2}
\, \yba^3 \lb{BETL}\\
\bt_{\lambda}(\yba,\lmb) &\equiv& \frac{d \lmb(t)}{dt} = \frac{3}{2\pi^2}
\, \lmb^2 + \frac{N_D}{ \pi^2} \, \lmb \, \yba^2
- \frac{N_D}{\pi^2} \, \yba^4 \;,
\lb{BETY}
\eea
where $\lmb$ and $\yba$ are the running couplings and
$t=\ln (\mu/\Lambda_1)$. The quantity $\mu$ is an energy scale
and $\Lambda_1$ is the cut-off of the one-loop $\bt$-function model.
In order to find the relations for the renormalized
couplings we have integrated these $\bt$-functions numerically
from the cut-off scale $\mu=\Lambda_1$
down to the physical scale $\mu=v_{R}$, with the identifications
$\yba(0)=y_0$, $\yba(t_R)=y_R$, where $t_R=\ln(v_R/\Lambda_1)$.
%
%
%
%
%
Since the simulations were carried out at $\lm=\infty$, we have chosen
$\lm_0 =100 \approx \infty$ and varied the starting value for $y_0$
in the interval $0 \leq y < 50$.
The solid and dotted curves in fig.~\ref{FLindn} correspond to two different
values of the ratio $v_R/\Lm_1$.
For the solid curve this ratio was adjusted such that
the agreement with the numerical results is best.
It is remarkable that the
shape of the curve is in reasonable agreement with our data.
The ratio $v_R/\Lm_1$ of the dotted curve is by a factor two smaller
than that of the solid curve. \\

Let us now turn to the question of the determination of upper bounds on
$m_{\sigma}$ and $m_F$.
Reading off an upper bound
at $m_{\sg}=0.7/a$, we find $m_{\sg}/v_R \aleq 3$  in region (I) of the
phase diagram and $m_{\sg}/v_R \aleq 4$ in regions (II) and (III).
The fermion mass  increases along the phase transition line (keeping
$am_{\sg}$ constant) until $y \approx 4$ where $\kp_c(y)$ becomes negative.
{}From then on it appears to remain roughly constant.
Also the field expectation value does not change much for
$y\ageq 4$ and we find $m_F/v_R=0.9$, 2.6 and 2.6 in regions
(I), (II) and (III) respectively. The observed saturation of the
renormalized Yukawa coupling indicates that we can read off the
upper bound for $y_R$ at $y \approx 4$. \\

%
\begin{figure}
\hspace*{0.6cm}
\centerline{
\fpsxsize=14.0cm
\fpsbox{Fdiscuss.ps}
}
\vspace*{0.7cm}
 \caption{ \noindent {\em Comparison of the results for
 the ratio $m_{\sigma}/v_R$ as a function
 of $m_F/v_R$ obtained in various models. The open symbols represent
 infinite volume results for the O(4) model ($N_D=0$, square)
 { \protect \cc{O4} }, the naive fermion model ($N_D=32$, open circle)
 { \protect \cc{BoDe91} } and the reduced staggered fermion model
 ($N_D=2$, diamond), whereas the full circles show finite volume
 ($6^3 12$) results obtained in the mirror fermion model ($N_D=2$)
 { \protect \cc{FrLi92} }.
 }}
\label{FlmRyR}
\end{figure}
It is interesting to compare our results for $\lm_R$
with those obtained in the mirror fermion model of ref.~\cite{Mo87},
which also describes two doublets of light fermions. Since the
mirror fermion model has a quite different bare action
with a different bare parameter space,
a comparison of the $y_R$ dependence
of $\lm_R$ at the same values of $a v_R$ would give
some idea to what extent these results are model dependent.
Such a comparison is shown in
fig.~\ref{FlmRyR} where we have plotted
the ratio $m_{\sg}/v_R$ against $m_F/v_R$ for various models.
The three diamonds represent some infinite volume results
in the reduced staggered fermion model which have
roughly the same cut-off, $a v_R=0.15$-$0.21$.
In contrast to our results,
the results in the mirror fermion model (full circles)
\cite{FrLi92} were not extrapolated to infinite volume.
The cut-off values on the $6^3 12$ lattice,
$a v_R =0.31$-$0.39$, are consistent with our results on the $6^3 24$
lattice (see table 2). It can be seen from fig.~\ref{FlmRyR} that
the points obtained from the mirror model
are within error bars on the same curve as our results. \\

We included in fig.~\ref{FlmRyR} also some infinite volume
results which were obtained in the pure O(4) model ($N_D=0$, square,
$a v_R=0.20(2)$)
\cc{O4} and in the naive fermion model ($N_D=32$, open circle,
$a v_R=0.20(3)$) \cite{BoDe91}.
It is seen that in the naive fermion model, where $m_F/v_R$
is relatively small, the fermion effect on $\lm_R$ is negligible, whereas
the fermions give a moderate
increase of $\lm_R$ in the model with two doublets. In both
models the maximal values for $y_R$ appear to
be close to the tree level unitarity bound given in eq.~(\eq{TUB}). \\

An interesting speculation has been that, even though the
Yukawa and quartic self-couplings are trivial, they might still be
sufficiently large as to give rise to
interesting non-perturbative effects, like
formation of a $\rh$ bound state \cc{RHO}.
We find, however, that the largest values of $\lm_R$ and $y_R$ we can obtain
are very close to their tree level unitarity bounds,
which suggests that they are still on the edge of the perturbative regime.
This indication is supported by the observation here and also
in ref.~\cite{BoDe91} for the 32 doublet model,
that the numerical results for the scalar propagator
can be very accurately described using one-loop renormalized perturbation
theory. It is therefore most likely that renormalized perturbation theory
gives a complete and accurate description of the physics of Yukawa models.
%
%
\section{Summary and discussion}
%
%
In this paper we have used `reduced' or `real' staggered fermions in
a high statistics investigation of a fermion-Higgs model.
The two staggered flavors are coupled to the O(4) Higgs
field, leading to a model with a single isospin doublet in the
scaling region. In a simulation with the Hybrid Monte Carlo
algorithm this number of doublets has to be
doubled. Since the species doubler degrees of freedom are used as
physical flavor-spin components, there is no redundancy in the fermion
field in this approach and the model can be very efficiently
simulated on large lattices. \\

The Yukawa coupling to the staggered flavors breaks the O(4)
symmetry and it requires two scalar field counterterms
to restore this symmetry in the
scaling region. Even without these counterterms,
this O(4) symmetry breaking has only a small effect on the
values of the renormalized couplings, at least in the physically
relevant scaling region of the
phase diagram where we have done our simulations. This follows from
our findings that a) the
deviations from the $1/L^2$ finite size effects are very small,
b) the Goldstone mass is much smaller than the Higgs mass
and c) that the one fermion loop estimate of the symmetry breaking
correction to the quartic coupling is found to be small.
The one fermion loop estimate for the symmetry breaking
effects is presumably reliable, because is gives a reasonable prediction
for the Goldstone mass. \\

The renormalized Yukawa coupling $y_R$ and the
renormalized quartic self-coupling $\lm_R$
were determined from the tree level relations $y_R=m_F/v_R$ and
$\lm_R=m_{\sg}^2/2v_R^2$, where $v_R=v/\sqrt{Z_{\pi}}$ is the renormalized
field expectation value. On a finite lattice the masses
$m_F$ and $m_{\sg}$ and the Goldstone
wave-function renormalization constant $Z_{\pi}$
were computed from the scalar and fermion propagators in momentum space.
The momentum dependence of the scalar propagators differs from
that of a free propagator and in particular on small lattices it is
important to include the one fermion loop self-energy to get a reliable
estimate for $m_{\sg}$ and $Z_{\pi}$.
In order to investigate the finite size
dependence of our results and
extrapolate them to the infinite volume, we have simulated
on a sequence of lattices ranging in size from $6^3 24$ to $16^3 24$.
The finite volume results for $m_{\sg}$, $v_R$ and $m_F$
show a $1/L^2$ dependence,
which was used to extrapolate these observables
to the infinite volume. \\

A comprehensive summary of our infinite volume results is shown
in fig.~\ref{FLindn},
where we have plotted the ratio $m_{\sg}/v_R=\sqrt{2 \lm_R}$ as a function
of $m_F/v_R=y_R$. The results were obtained in three different
regions of the phase diagram (c.f. fig.~\ref{Fphased}) and always at
$\lm=\infty$. The ratio $m_{\sg}/v_R$ increases a little bit for
increasing $y_R$ and becomes at large $y_R$
slightly larger than the tree level unitarity
bound. The maximal value of the Yukawa coupling is roughly equal to its tree
level unitarity bound. This indicates that the couplings are not very strong.
Further evidence for the perturbative nature
of the renormalized couplings comes from the fact that
the numerical results for the scalar field propagators can be nicely described
by taking into account the one fermion loop contribution to the self-energy.
We find even for the points in region (III)
an almost perfect agreement over the full range of momenta. Furthermore
fig.~\ref{FLindn} shows that the numerical values
for $m_{\sg}/v_R$ and $m_F/v_R$ are in qualitative agreement
with a simple model based
on extending the one-loop $\bt$-function
to infinite couplings (solid curve). In conclusion we can say that all
our results are consistent with the triviality
scenario and there is no evidence for a strong coupling sector. \\

Our results are obtained in a particular regularization of the SU(2)
fermion-Higgs model: using the lattice regularization with
reduced staggered fermions and two isospin doublets in the scaling
region. One would like to see to what extent the results are
model dependent. Therefore we have compared our results with those
of ref.~\cc{FrLi92} which uses mirror
fermions with two doublets in the scaling region. Looking
at the $y_R$ dependence of $\lm_R$ at roughly the same value of the
cut-off in all models, we find consistent results within error
bars, c.f.~fig.~\ref{FlmRyR}. \\

The main result of our investigation therefore is that the renormalized
quartic and Yukawa couplings cannot be strong, unless
the cut-off is unacceptably low. In fact we find that
one-loop renormalized perturbation theory is
applicable even for the maximally strong couplings.
More quantitatively we find for the upper bounds on the masses
of the Higgs particle and the heavy fermion
in our model, $m_{\sg} \aleq 4v_R$ and $m_F \aleq 2.6v_R$ for cut-off
values $\Lambda \ageq 2.5 \; m_{\sg,F}$. From experience in the O(4) model
with modifications in the regularization (by including dimension
six operators \cite{HeNe92} or using improved actions \cite{GoKa92})
we expect that these numbers for the upper bounds may
be stretched by perhaps 20-30\%. Larger changes have been recently
found with the Pauli-Villars regularization scheme in the large $N$ limit
\cc{Ku}.
\subsubsection*{Acknowledgements}
We thank J.~Jers\'ak for useful discussions.
The numerical calculations were performed on the CRAY Y-MP4/464
at SARA, Amsterdam, on the S600 at RWTH Aachen
and on the CRAY Y-MP/832 at HLRZ J\"ulich.
This research was supported by the ``Stichting voor
Fun\-da\-men\-teel On\-der\-zoek der Materie (FOM)''
and by the ``Stichting Nationale Computer Faciliteiten (NCF)''.
%
%

%
%
\end{document}